\documentclass[aip,apl,reprint,superscriptaddress%,groupedaddress
]{revtex4-1}
\usepackage{graphicx}
\usepackage{amsmath,amsfonts,amssymb}
\usepackage{xcolor}
\usepackage{epstopdf}

\newcommand{\red}[1]{\textcolor{red}{#1}} %added by authors

\draft % marks overfull lines with a black rule on the right

\begin{document}

%\preprint{}

%\title{Defects, band edges, and optically-determined mobility in the p-type transparent conductor CuI}
\title{Assessing the defect tolerance of kesterite-inspired solar absorbers}



\author{Andrea Crovetto}
\email[]{Electronic mail: andrea.crovetto@helmholtz-berlin.de}
%\thanks{}
\affiliation{SurfCat, DTU Physics, Technical University of Denmark, DK-2800 Kgs. Lyngby, Denmark}
\affiliation{Department of Structure and Dynamics of Energy Materials, Helmholtz-Zentrum Berlin f\"ur Materialien und Energie GmbH, 14109 Berlin, Germany}

\author{Sunghyun Kim}
%\thanks{}
\affiliation{Department of Materials, Imperial College London, London SW7 2AZ, United Kingdom}

\author{Moritz Fischer}
%\thanks{}
\affiliation{DTU Fotonik, Technical University of Denmark, DK-2800 Kgs. Lyngby, Denmark}
\affiliation{Center for Nanostructured Graphene (CNG), Technical University of Denmark, DK-2800 Kgs. Lyngby, Denmark}

\author{Nicolas Stenger}
%\thanks{}
\affiliation{DTU Fotonik, Technical University of Denmark, DK-2800 Kgs. Lyngby, Denmark}
\affiliation{Center for Nanostructured Graphene (CNG), Technical University of Denmark, DK-2800 Kgs. Lyngby, Denmark}

\author{Aron Walsh}
%\thanks{}
\affiliation{Department of Materials, Imperial College London, London SW7 2AZ, United Kingdom}
\affiliation{Department of Materials Science and Engineering, Yonsei University, Seoul 03722, Korea}

\author{Ib Chorkendorff}
%\thanks{}
\affiliation{SurfCat, DTU Physics, Technical University of Denmark, DK-2800 Kgs. Lyngby, Denmark}

\author{Peter C. K. Vesborg}
%\thanks{}
\affiliation{SurfCat, DTU Physics, Technical University of Denmark, DK-2800 Kgs. Lyngby, Denmark}

\begin{abstract}
Various thin-film I$_2$-II-IV-VI$_4$ photovoltaic absorbers derived from kesterite Cu$_2$ZnSn(S,Se)$_4$ have been synthesized, characterized, and theoretically investigated in the past few years. The availability of this homogeneous materials dataset is an opportunity to examine trends in their defect properties and identify criteria to find new defect-tolerant materials in this vast chemical space.
We find that substitutions on the Zn site lead to a smooth decrease in band tailing as the ionic radius of the substituting cation increases. Unfortunately, this substitution strategy does not ensure the suppression of deeper defects and non-radiative recombination.
Trends across the full dataset suggest that Gaussian and Urbach band tails in kesterite-inspired semiconductors are two separate phenomena caused by two different antisite defect types. Deep Urbach tails are correlated with the calculated band gap narrowing caused by the (2I$_\mathrm{II}$+IV$_\mathrm{II}$) defect cluster. Shallow Gaussian tails are correlated with the energy difference between the kesterite and stannite polymorphs, which points to the role of (I$_\mathrm{II}$+II$_\mathrm{I}$) defect clusters involving Group IB and Group IIB atoms swapping across \textit{different} cation planes. This finding can explain why \textit{in-plane} cation disorder and band tailing are uncorrelated in kesterites. Our results provide quantitative criteria for discovering new kesterite-inspired photovoltaic materials with low band tailing.
%These out-of-plane clusters are less abundant than the in-plane (I$_\mathrm{II}$+II$_\mathrm{I}$) clusters, which are responsible for the experimentally quantifiable cation disorder and for an overall band gap shift.
%This subtle difference can explain why various studies have found a lack of correlation between cation disorder and band tails in kesterite and can guide future design of kesterite-inspired photovoltaic materials.
\end{abstract}


\maketitle %\maketitle must follow title, authors, abstract and \pacs

%%%MAIN TEXT%%%%
\section{Introduction}
Progress in the photovoltaic efficiency of kesterite Cu$_2$ZnSn(S,Se)$_4$ solar cells has been minimal after reaching the 12.6\% efficiency mark in 2013.~\cite{Giraldo2019a} While interface-related issues may be solved by an appropriate choice of contact layers,~\cite{Crovetto2018a,Crovetto2017} the unforgiving native defect chemistry of Cu$_2$ZnSn(S,Se)$_4$ kesterites~\cite{Chen2013,Kim2019} implies that bulk-related issues may be more difficult to overcome. In fact, fast non-radiative recombination~\cite{Hages2017} and band tailing~\cite{Rey2018} are observed in Cu$_2$ZnSn(S,Se)$_4$ regardless of growth technique, stoichiometry and chemical potentials during growth. Some of these concerns were recently quantified by a combination of first-principles defect calculations and device simulation, which led to the estimation of an upper efficiency limit of only 20-21\% for both Cu$_2$ZnSnS$_4$ and Cu$_2$ZnSnSe$_4$, as opposed to the upper limit of 32\% for a defect-free absorber of the same band gap.~\cite{Kim2020} The efficiency limitation was derived by calculating the open circuit voltage loss associated with non-radiative recombination through various native defects. 
Since tail states were not considered in the simulation, the realistic efficiency potential of Cu$_2$ZnSn(S,Se)$_4$ solar cells is probably even lower.~\cite{Gokmen2014}

A possible strategy to mitigate the efficiency losses due to non-radiative recombination and band tailing is to perform isoelectronic element substitutions on the Cu$_2$ZnSn(S,Se)$_4$ template in the hope to obtain a more defect-tolerant material than the original.~\cite{Wang2014a,Shin2017,Giraldo2019a,Pandey2018} A popular approach has been to partially substitute certain cations with small amounts of other cations (e.g. Sn with Ge and Cu with Li).~\cite{Giraldo2017,Cabas-Vidani2018} However, the chemical trends in the defect tolerance of these kesterite-inspired compounds may be easier to discern by considering the fully substituted materials, since non-linear alloying effects are avoided and comparison between experiment and theory is more straightforward. Some fully-substituted, kesterite-inspired absorbers have received considerable attention and have achieved efficiencies above 5\%: Ag$_2$ZnSnSe$_4$,~\cite{Gershon2016a} Cu$_2$ZnGeSe$_4$,~\cite{Sahayaraj2017} Cu$_2$CdSnS$_4$,~\cite{Hadke2019} and Cu$_2$BaSn(S,Se)$_4$ (CBTSSe).~\cite{Shin2017a} Interestingly, solar cells based on the pure sulfides Cu$_2$SrSnS$_4$ (CSTS) and Cu$_2$BaSnS$_4$ (CBTS) were also reported recently.~\cite{Ge2017,Crovetto2019a} This gives access to a series of four Cu$_2$-II-Sn-S$_4$ (CXTS) sulfides (X = Zn, Cd, Sr, Ba). The four X$^{2+}$ cations from Groups IIA and IIB cover a wide range of ionic radii (Table~\ref{tab:ionic_radii}), which are a classical descriptor of defect formation energies in semiconductors.\cite{Hautier2011} Thus, studying chemical trends in the CXTS series could give insights into the fundamental mechanisms behind strong band tailing and fast non-radiative recombination in kesterite absorbers.

\begin{figure*}[t!]
\centering%
\includegraphics[width=0.9\textwidth]{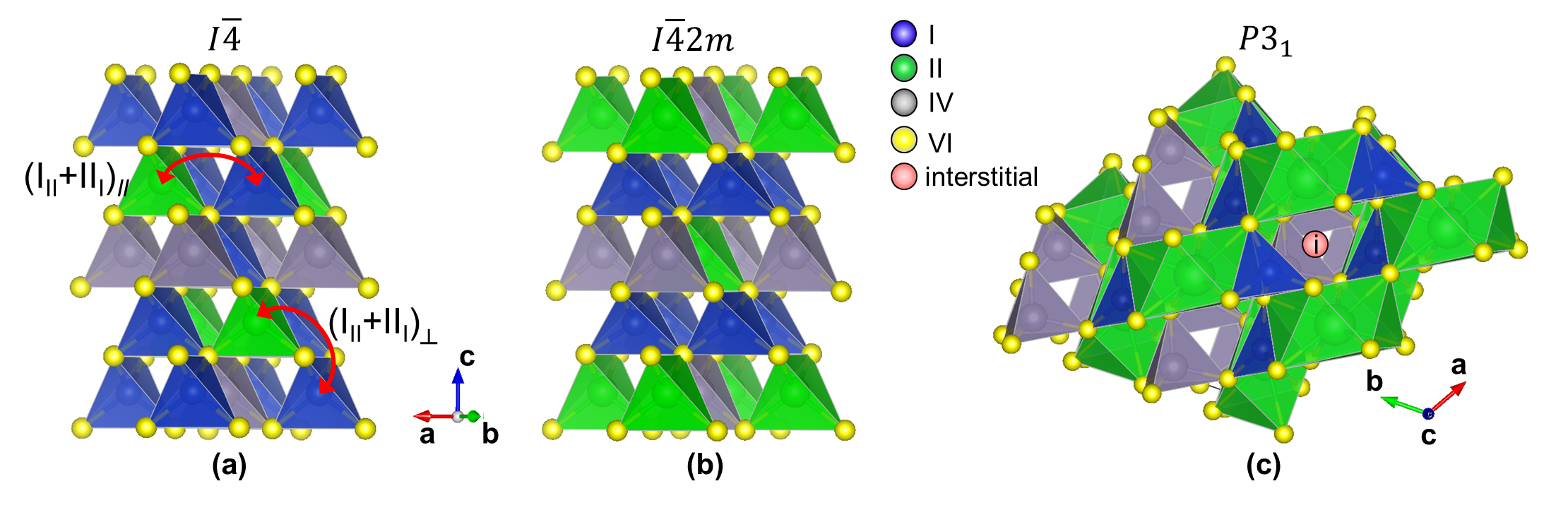}
\caption{(a) The $I\bar{4}$ (kesterite) structure. (b) The $I\bar{4}2m$ (stannite) structure. (c) The trigonal $P3_1$ structure. The difference between the (I$_\mathrm{II}$+II$_\mathrm{I}$)$_{\parallel}$ defect cluster and the (I$_\mathrm{II}$+II$_\mathrm{I}$)$_{\perp}$ defect cluster is visualized in (a). Notice that a kesterite structure with one (I$_\mathrm{II}$+II$_\mathrm{I}$)$_{\perp}$ cluster per 8-atom unit cell is equivalent to the stannite structure. In (c), we have drawn a hypothetical interstitial defect located in the "cage" between two square antiprisms formed by the anions surrounding the Group II cation.}
\label{fig:structures}
\end{figure*}

Cu$_2$ZnSnS$_4$ (CZTS) and Cu$_2$CdSnS$_4$ (CCTS) have similar zincblende-derived, tetrahedrally-coordinated structures ($I\bar{4}$ kesterite and $I\bar{4}2m$ stannite respectively) with similar band gaps (1.4-1.5~eV). Cu$_2$SrSnS$_4$ (CSTS) and Cu$_2$BaSnS$_4$ (CBTS) have wider band gaps ($\sim$2.0~eV) and crystallize in the trigonal $P3_1$ structure where Cu and Sn are tetrahedrally coordinated but the larger II$^{2+}$ cation is 8-fold coordinated (Fig.~\ref{fig:structures}).

\begin{table}[t!]
\centering
\setlength\tabcolsep{4pt}
\begin{tabular}{l c c c c c}
\hline
 & Coord. & Ionic &  & Crystal & \\
 Cation & number & radius (\AA)  & compound & structure & $\varepsilon_\mathrm{s}$  \\
\hline
Cu$^{+}$ & 4 & 0.60 & & \\
Sn$^{4+}$ & 4 & 0.55  & & \\
Zn$^{2+}$ & 4 & 0.60  & Cu$_2$ZnSnS$_4$ & $I\bar{4}$ & 6.8  \\
" & " & "  & Cu$_2$ZnSnSe$_4$ & $I\bar{4}$ & 8.6  \\
Cd$^{2+}$ & 4 & 0.78 & Cu$_2$CdSnS$_4$ & $I\bar{4}2m$ &   \\
Sr$^{2+}$ & 8 & 1.26 & Cu$_2$SrSnS$_4$ & $P3_1$  & 6.1 \\
Ba$^{2+}$ & 8 & 1.42 & Cu$_2$BaSnS$_4$ & $P3_1$ & 6.1  \\
Ag$^{+}$ & 4 & 1.00 & Ag$_2$ZnSnSe$_4$ & $I\bar{4}$ & 12.6  \\
Ge$^{4+}$ & 4 & 0.39 & Cu$_2$ZnGeSe$_4$ & $I\bar{4}$ & \\
\hline
\end{tabular}
\caption{Selected properties of the kesterite-inspired compounds investigated in this study and of their constituent cations. As a measure of the cation size, we use Shannon's effective ionic radii.~\cite{Shannon1976a} $\varepsilon_\mathrm{s}$ is the static dielectric constant (relative permittivity) responsible for screening electrostatic potential fluctuations. Values for CBTS and CSTS were calculated in this study. Values for the other materials are taken from Refs.~\citenum{Persson2010,Gershon2017b}.}
\label{tab:ionic_radii}
\end{table}

As mentioned above, an interesting feature of the CXTS series is the progressive increase of ionic radius of the X$^{2+}$ cation from Zn$^{2+}$ (0.60 \AA) to Ba$^{2+}$ (1.42 \AA). The increasing ionic radius corresponds to an increasing size mismatch between the II$^{2+}$ cation and both the Cu$^{+}$ and Sn$^{4+}$ cations (Table~\ref{tab:ionic_radii}). Thus, classical intuition suggests that substituting Zn with the largest II$^{2+}$ cations (Sr and Ba) presents an opportunity to reduce the concentration of various antisite defects, due to the high energetic cost of forming antisite defects between highly size-mismatched cations.~\cite{Hong2016} Despite this potential advantage, the record efficiencies of CBTS and CSTS solar cells (2.0\% and 0.6\% respectively)~\cite{Ge2017,Crovetto2019a} are much lower than those of CZTS and CCTS solar cells (11.0\% and 8.0\% respectively).~\cite{Hadke2019,Yan2018b}

In the first part of this work, we investigate chemical trends in the defect properties of the CXTS series by a combination of experimental techniques and first-principles calculations. Since defect spectroscopy measurements are not available in the literature for CBTS and CSTS, we start by performing temperature- and excitation dependent photoluminescence (PL) measurements on these materials and assign their shallow defect transitions to Cu vacancies and Cu interstitials by matching their measured ionization energies to the calculated charge transition levels. Across the whole Cu$_2$-II-Sn-S$_4$ series, we find that increasing the size of the II$^{2+}$ cation leads to a smooth decrease in band tailing.
In the second part of this work, we expand our investigation to a wider range of kesterite-inspired materials involving the substitution of other cations and anions. Analysis of their band tail trends suggests that Gaussian and Urbach tails in this class of materials have different chemical origins. Based on correlations between measured band tail parameters and calculated quantities, we tentatively assign Urbach tails to the (2I$_\mathrm{II}$+IV$_\mathrm{I}$) defect clusters and Gaussian tails to the (I$_\mathrm{II}$+II$_\mathrm{I}$) defect cluster involving cations on different cationic planes. Finally, we conclude that non-radiative recombination and band tailing are largely decoupled from one another in kesterite-inspired materials. For example, CBTS and CSTS have more non-radiative losses than CZTS, although they exhibit significantly less band tailing.

\section{Experimental and computational details}
CBTS and CSTS films on Mo-coated soda lime glass (SLG) were synthesized by sulfurization of oxide precursor films deposited by reactive sputtering. CZTS films on Mo-coated SLG were synthesized by sulfurization of sulfur-deficient CZTS films deposited by pulsed laser deposition. Details of the growth processes are available in previous publications.~\cite{Crovetto2019a,Crovetto2019b,Cazzaniga2017} The elemental composition and sulfurization conditions of the films characterized in this work are those that gave the highest-efficiency cells in the previous studies.~\cite{Crovetto2019a,Crovetto2019b,Cazzaniga2017} For all compounds, the bulk composition is Cu-poor and II-rich, and the sulfurization temperature is in the 520-560$^\circ$C range. For photoluminescence spectroscopy, the films were measured as-sulfurized without top contact layers. For external quantum efficiency measurements, a CdS/ZnO/ITO top contact was deposited by chemical bath deposition/RF sputtering/RF sputtering respectively.

Temperature-dependent and excitation intensity-dependent photoluminescence (PL) measurements were performed with a customized scanning microscopy setup based on a Nikon Eclipse Ti-U inverted microscope and a continuous wave (CW) 523~nm laser. The sample was placed inside a temperature controlled stage (HFS600, Linkam Scientific Instruments). Using a beam splitter, laser light was focused on the sample by a 10x objective lens and PL emission was collected by the same objective. The spot size was $\sim$11~$\mu$m and the excitation intensity was $\sim$400~mW/mm$^2$ for the temperature-dependent measurements. PL emission was filtered by a 550 nm long pass filter and directed to a spectrometer (Shamrock 303i, Andor) equipped with an electronically cooled CCD detector through a 250~$\mu$m input slit. 
Additional PL spectra over a larger area (about 1 mm$^2$) were used to quantify the relative PL intensity across various materials. An Accent RPM2000 system with 405~nm continuous-wave excitation laser at power density 500~mW/mm$^2$ was used for this purpose.
The external quantum efficiency (EQE) of the solar cells was measured using a PV Measurements QEXL setup calibrated with a reference Si photodiode.

Defect formation was probed from first-principles using the supercell approach. 
Calculations were performed based on density functional theory (DFT)~\cite{Hohenberg1964, Kohn1965} using the projector-augmented wave (PAW) method \cite{Blochl1994} and the hybrid exchange-correlation functional of Heyd-Scuseria-Ernzerhof (HSE06)~\cite{Heyd2003} as implemented in VASP.~\cite{Kresse1999}
The wave functions were expanded in plane waves up to an energy cutoff of 380~eV.
The \textit{k}-points were sampled according to a Monkhorst-Pack $k$-mesh~\cite{Monkhorst1976} with a grid spacing less than $2\pi\times$0.03 \AA$^{-1}$ for Brillouin zone integration.
The atomic coordinates were relaxed until the forces were less than 0.01~eV/\AA. %
The lattice vectors were optimized until residual stress was below 0.5~kbar.
To eliminate the spurious electrostatic interactions between charged defects,
finite size corrections~\cite{Freysoldt2009,Kumagai2014} were employed.

\section{Results and discussion}
\subsection{Photoluminescence features of Cu$_2$-II-Sn-S$_4$ compounds}
PL features of CZTS have been extensively discussed in the literature~\cite{Levcenko2016,Gershon2013,Grossberg2013,Tanaka2014,Grossberg2012} and were recently summarized.~\cite{Grossberg2019} Briefly, the PL peak of CZTS thin films at room temperature (RT) is broad and significantly red-shifted with respect to the band gap (FWHM and Stokes shift of more than 150~meV). Analysis of temperature-dependent and excitation power-dependent PL indicates strong spatial fluctuations in the band gap or the electrostatic potential of CZTS, or both. There is some consensus that the PL peak usually observed at RT can be attributed either to a band-to-impurity transition, also known as free-to-bound (FB),~\cite{Levcenko2016} or a tail-to-impurity (TI) transition.~\cite{Grossberg2012} Assignment to either transition may depend on how tail states are defined and on the specific samples being characterized. The impurity in these transitions is usually attributed to the Cu$_\mathrm{Zn}$ acceptor or the Zn$_\mathrm{Cu}$ donor, both expected to be abundant and with compatible ionization energies.~\cite{Chen2013}
The pure selenide CZTSe has qualitatively similar PL features to the pure sulfide CZTS, although its room-temperature Stokes shift and peak width are generally smaller than in the pure sulfide, and the defect involved in the FB transition is generally shallower. In fact, some authors argue that impurities in CZTSe are too shallow to be distinguishable from the overall tail states, at least at RT.~\cite{Rey2018} Band-to-band (BB) transitions are rarely detected in CZTS and CZTSe by PL, although they have been reported in both compounds under high excitation intensity at RT.~\cite{Chagarov2016,Grossberg2013,Tanaka2014}

PL features of CCTS have also been discussed in the literature, although less extensively than for CZTS.~\cite{Hadke2019,Pilvet2017} Similarly to the case of CZTS, the main room-temperature PL peak of CCTS is broad, strongly Stokes-shifted, and attributed to a FB transition. However, a narrower band-to-band peak is clearly observed in high-quality films.~\cite{Hadke2019} PL features of CBTS and CSTS have not been analyzed in detail,~\cite{Ge2017b,Crovetto2019a,Crovetto2019b} so in the the next two sections we will discuss temperature- and excitation dependent PL measurements on our own CBTS and CSTS films.

\begin{figure*}[t!]
\centering%
\includegraphics[width=0.8\textwidth]{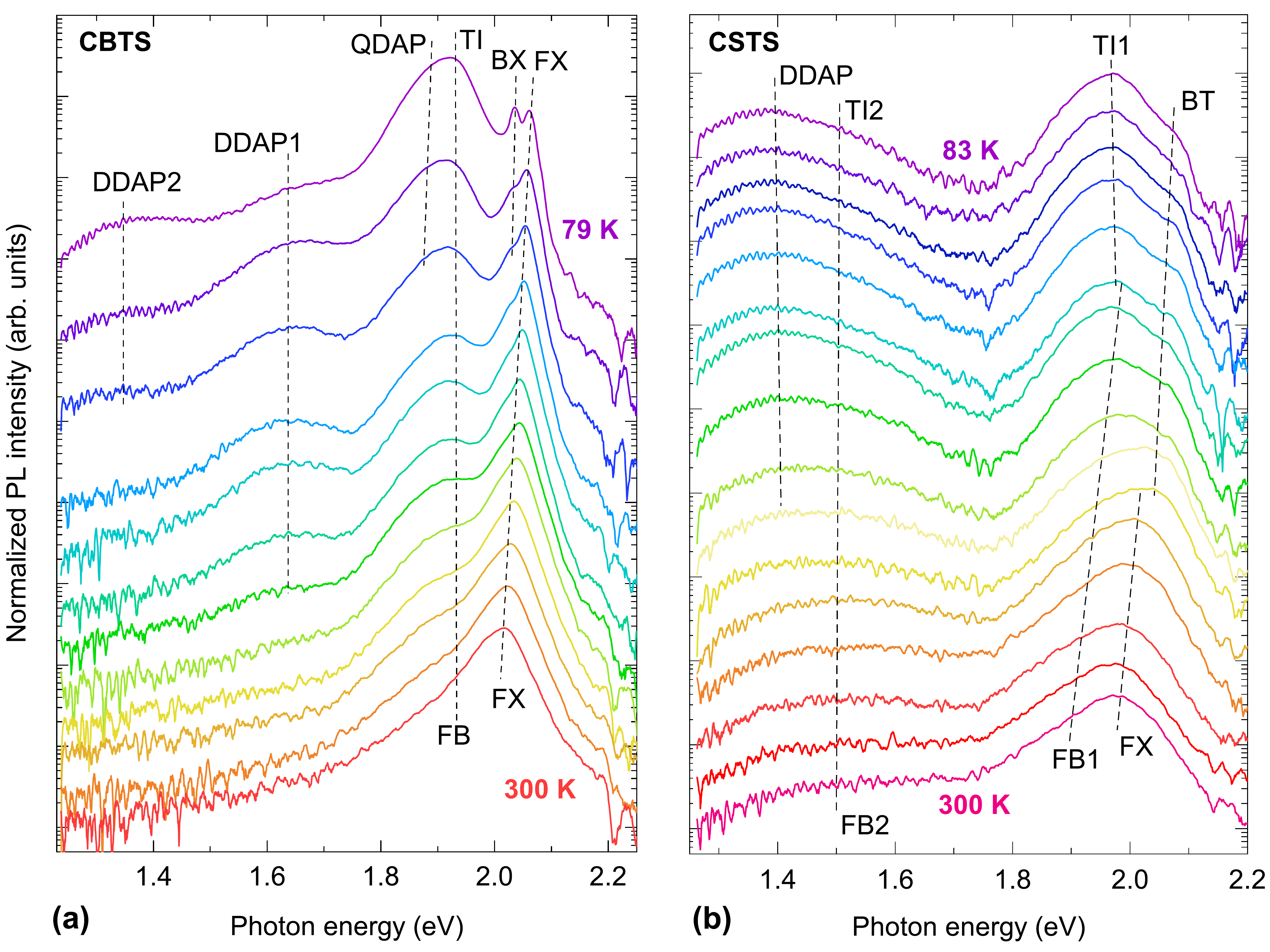}
\caption{Temperature-dependent PL spectra of CBTS (a) and CSTS (b). The maximum and minimum temperature are indicated. The dashed lines are guides to the eye for following the evolution of various PL transitions as a function of temperature. The transitions are labeled using the abbreviations introduced in the main text.}
\label{fig:T_dep}
\end{figure*}

\subsubsection{Photoluminescence of CBTS}
PL spectra of CBTS are complex, with at least five distinct peaks recognizable at 79~K (Fig.~\ref{fig:T_dep}(a)).
We will propose a possible interpretation of PL features and related defects based on the data in Fig.~S1, Supporting Information using the interpretation rules listed in Ref.~\citenum{Siebentritt2006}. The results will be summarized in Table~\ref{tab:defects_CBTS}. Plots of the integrated peak areas ($I_\mathrm{PL}$), peak widths (FWHM) and peak positions ($E_\mathrm{peak}$) as a function of temperature $T$ and excitation intensity $I_\mathrm{ex}$ are considered.~\cite{Siebentritt2006} Additional supporting data is provided by the exponent $k$ of the power law $I_\mathrm{PL} \propto I^k_\mathrm{ex}$ (Fig.~S1(d)) and the activation energy $E_\mathrm{act}$ of shallow defect levels obtained by least-squares fitting of Arrhenius plots of $I_\mathrm{PL}$ (Fig.~S1(b)). Finally, the energy shift between a pair of peaks, or between a peak and the band gap energy can be useful for determining defect levels. All reported values are based on least-squares peak fitting using Gaussian functions with three fitting parameters (peak area, FWHM, and position).

Two narrow peaks at 2.038~eV and 2.065~eV are observed at 79~K, with peak positions independent of excitation intensity (Fig.~S1(e)). Their narrow linewidth is typical of excitonic transitions~\cite{Siebentritt2006} using the model of a series of Wannier excitons with S orbitals. Assignment to the first and second free exciton of CBTS is excluded because the higher-energy peak has roughly the same intensity of the lower-energy peak. Instead, we assign the peak at 2.065~eV to a free exciton (FX) and the peak at 2.038~eV to a bound exciton (BX) similarly to the case of Cu-rich CuGaSe$_2$.~\cite{Bauknecht2000} The possible origin of the defect involved in the BX peak will be discussed later.
The FX peak broadens and red-shifts with increasing temperature, with a position of 2.022~eV at RT (Fig.~S1(a)). On the other hand, the BX peak becomes difficult to distinguish already at 100~K due to thermal ionization of the involved defect. Band-to-band (BB) recombination often takes over excitonic recombination in inorganic semiconductors with a low exciton binding energy $E_\mathrm{b}$ and a high dielectric constant $\varepsilon_s$ as the temperature is increased.~\cite{Siebentritt2006,Larsen2011} This change in recombination mechanism can be detected by the shift of the FX/BB peak versus temperature. The net effect of two phenomena determines the thermal peak shift: 1) the temperature-dependent band gap change, and 2) a blue shift given by $E_\mathrm{b}$ when BB recombination takes over FX recombination. We find that the FX peak of CBTS red-shifts by 43~meV between 79~K \red{and} RT, in good agreement with the known temperature coefficient of the CBTS band gap.~\cite{Ge2017} We estimate the exciton binding energy of CBTS as $\simeq 65$~meV from the hydrogen model using our calculated dielectric constant of 6.1 and average electron- and hole effective masses of 0.22~$m_0$ and 0.92~$m_0$, respectively~\cite{Zhu2017,Pandey2018} ($m_\mathrm{0}$ is the electron rest mass). A blue shift of 65~meV would be quite substantial, yet it is not observed in Fig.~S1(a).

%However, the 2.065~eV peak is found to red shift with increasing $T$ and is centered at 2.022~eV at RT, which is compatible with the known temperature coefficient of the CBTS band gap.~\cite{Ge2017}
%However, we find a rather low static dielectric constant of 6.1 from first-principles calculations and a rather high exciton binding energy of $\simeq 65$~meV from the hydrogen model using 0.22~$m_0$ and 0.92~$m_0$ as the average electron- and hole effective masses in CBTS~\cite{Zhu2017,Pandey2018} ($m_\mathrm{0}$ is the electron rest mass). If a BB transition were dominant at RT, the position of the low-temperature 2.065~eV peak would change according to the net effect of two opposing tendencies, i.e., a red-shift due to thermal narrowing of the band gap and a blue-shift by an energy equal to $E_\mathrm{b}$. However, the 2.065~eV peak is found to red shift with increasing $T$ and is centered at 2.022~eV at RT, which is compatible with the known temperature coefficient of the CBTS band gap.~\cite{Ge2017} 

Hence, we conclude that the dominant PL peak of CBTS at RT is a (broadened) excitonic transition, as in layered halide perovskites and in many organic semiconductors.~\cite{Ishihara1990,Wurfel2010} The relatively low $\varepsilon_s$ and relatively high $E_\mathrm{b}$ in CBTS may be responsible for the persistence of excitonic transitions up to RT. This conclusion can explain the characteristic dip in the RT absorption coefficient of CBTS just above the band gap, which is typical of excitonic absorption~\cite{Ishihara1990} and has been observed in nearly all previous studies of CBTS.~\cite{Ge2017,Ge2017b,Crovetto2019b,Shin2017a} If the RT absorption onset is due to excitonic absorption, band gap extraction by means of Tauc plots is not justified and leads to underestimation of the band gap. Fitting the absorption coefficient with an Elliott function is a more appropriate method to extract the band gap in the presence of excitonic absorption.~\cite{Elliott1957} Based on the value of $E_\mathrm{b}$ estimated above, we suggest that the RT band gap of CBTS is $\sim$70-100~meV higher than the values previously estimated using Tauc plots.~\cite{Ge2017,Ge2017b,Crovetto2019b,Shin2017a}

The peak at $\sim$1.9~eV is the dominant PL feature at 79~K but it quenches with increasing temperature. At RT, it is merely detectable as a shoulder of the dominant FX peak. The Arrhenius plot in Fig.~S1(b) shows that the 1.9~eV peak quenches quickly between 79~K and 120~K, then it is approximately constant in intensity between 120~K and 200~K, and it finally quenches again above 200~K. This behavior cannot be easily explained for a single transition, so we conclude that the 1.9~eV peak is the convolution of two different peaks and we fit each of them with a Gaussian function up to $\sim$120~K. In this low-temperature range, both peaks red-shift with increasing temperature to a slightly larger amount than expected from thermal band gap narrowing. The position of the higher-energy peak is independent of excitation intensity, whereas the lower-energy peak blue-shifts by $\sim$10~meV/decade (Fig. S1(e)), suggesting that band tails exist in this temperature range due to band edge fluctuations.~\cite{Siebentritt2006} Thus, we identify the lower-energy peak as a quasi-donor-acceptor-pair (QDAP) peak and the higher energy peak as a tail-to-impurity (TI) transition which becomes dominant when the shallow defect of the QDAP transition becomes thermally ionized. The shallow defect level can be extracted from the TI$-$QDAP offset, which is $36 \pm 7$~meV in the temperature range where both peaks are detected (Fig. S1(a)). This value should is in good agreement with the activation energy of the QDAP peak, which is estimated as $45 \pm 5$~meV from a single exponential Arrhenius fit ($I_\mathrm{PL} = I_0 \exp(E_\mathrm{act}/k_\mathrm{B}T)$ in Fig. S1(b)). As the temperature increases, the TI peak blue-shifts in the 175-225~K range but does not shift further at higher temperatures. This behavior is very similar to the temperature evolution of the QDAP/TI peak in Cu-poor CuGaSe$_2$~\cite{Larsen2011} and can be explained by flattening of the band edge fluctuations in the region where the blue shift occurs. At temperatures above $\sim$225~K, band tails can be considered negligible so tail-to-impurity recombination effectively turns into band-to-impurity recombination, which is usually labeled as free-to-bound (FB) recombination. The mechanism responsible for band tail flattening at intermediate temperatures can be hypothesized by observing that the FWHM of PL peaks in CBTS generally increases with excitation intensity at low temperatures (Fig. S1(c)). If free carrier screening was the dominant mechanism the FWHM would decrease instead,~\cite{Rey2018,Lang2017} so tail state filling by photocarriers is a more likely mechanism. The impurity involved in the TI/FB transition is the same as the deep defect involved in the QDAP transition. This defect level can be extracted from the offset between the band gap energy (estimated as FX$+E_\mathrm{b}$) and the FB peak position at temperatures $>$225~K, where band edge fluctuations are flattened and thus do not influence the FB peak energy. This offset is $136 \pm 8$~meV, in good agreement with the activation energy of $144 \pm 9$~meV (Fig.~S1(b)) obtained by fitting the thermal quenching of the TI/FB peak with the equation

\begin{equation}
I_\mathrm{PL} = \frac{I_0}{1+aT^{3/2}\exp(-E_\mathrm{act}/k_\mathrm{B}T)}
\label{eq:FB}
\end{equation}
where $I_0$ is a constant, $a$ is a rate parameter, $E_\mathrm{act}$ is the activation energy, and $k_\mathrm{B}T$ is the thermal energy.

Two additional low-intensity peaks can be detected at around 1.65~eV and 1.35~eV at low temperatures. They cannot simply be phonon replicas of the TI/FB peak because they would require much higher phonon wavenumbers than the ones found experimentally in CBTS (below 400~cm$^{-1}$).~\cite{Ge2017,Crovetto2020b} Thus, these peaks must be related to other radiative transitions in CBTS involving deeper defects.
Small peak shifts versus excitation intensity are difficult to determine reliably for such low-intensity peaks.
%cannot be determined reliably because the fitted peak position of these low-intensity peaks is highly dependent on small changes of the fitted QDAP peak, or alternatively on the chosen background if they are fitted independently from the QDAP peak.
However, both peaks red-shift with increasing temperature similarly to the QDAP and TI peaks (Fig.~S1(a)), they exhibit rather small $k$ coefficients (Fig.~S1(d)), and their activation energies are much lower than those expected for such deep defects (Fig.~S1(b)). Therefore, we identify both peaks as QDAP transitions between a rather shallow defect (corresponding to the observed activation energy) and a deeper defect which cannot be thermally ionized at these temperatures. Since a deep defect is involved, we label the $\sim$1.65~eV and $\sim$1.35~eV peaks as DDAP1 and DDAP2 respectively, meaning deep DAP transitions. The activation energy of the DDAP1 peak, extracted with Eq.~\ref{eq:FB}, is $129 \pm 55$~meV, suggesting that the shallow defect involved in the DDPA1 transition may be the same defect that is also responsible for the TI/FB transition. Due to the limited available temperature range, the intensity of the DDAP2 peak is simply fitted with the $I_\mathrm{PL} = a \exp(E_\mathrm{act}/k_\mathrm{B}T)$ equation, yielding $E_\mathrm{act,DDAP2} = 39 \pm 5$~meV. This value is similar to the activation energy of the shallower defect involved in the QDAP transition, again suggesting the same chemical origin for both defects. The ionization energy $ E_\mathrm{i}$ of the deep defects involved in the two DDAP transitions can be estimated as
\begin{equation}
E_\mathrm{i} = \mathrm{FX} + E_\mathrm{b} - \mathrm{DDAP_x} - E_\mathrm{act,x} - 2\Gamma
\label{eq:deep_defects}
\end{equation}
where (FX$+E_\mathrm{b}$) is the low-temperature band gap, DDAP$_\mathrm{x}$ is the DDAP peak position, $E_\mathrm{act,x}$ is the activation energy of its shallower defect (as determined above) and $\Gamma$ is the average tail depth at one of the band edges. Estimating $2\Gamma \sim 30$~meV based on the blue shift of the TI/FB peak at intermediate temperatures (Fig.~S1(a)), ionization energies of $300 \pm 50$~meV and $700 \pm 50$~meV are derived for the deeper defects of the DDAP1 and DDAP2 transitions, respectively.

In conclusion, we have identified two shallow defects of opposite type (donor and acceptor) with ionization energies of about 40~meV and 135~meV. PL measurements alone cannot establish which is a donor and which is an acceptor.
%It cannot be established which one is a donor and which one is an acceptor on the basis of PL measurements alone.
We have also identified two deeper defects with $\sim$300~meV ionization energy (same type as the 40~meV defect) and $\sim$700~meV ionization energy (same type as the 135~meV defect). These results are summarized in Table~\ref{tab:defects_CBTS}. Band edge fluctuations exist in CBTS at low temperatures, but they are not sufficiently strong to dissociate excitons. The related tail states are filled by photocarriers at temperatures above 200~K.

\begin{table}[t!]
\centering
\begin{tabular}{l l c c }
\hline
 \multicolumn{2}{l}{CBTS: Experiment} & \multicolumn{2}{l}{CBTS: Theory} \\
Exp. quantity  & $E_\mathrm{i}$ (meV) & Defect & $E_\mathrm{t}$ (meV)  \\
\hline
$E_\mathrm{act,FB}$ & $144 \pm 9$ & Cu$_\mathrm{i}$(D1) & 118 \\
$E_\mathrm{g}(>225$~K)$-$FB & $136 \pm 8$ &  '' & ''  \\
$E_\mathrm{act,DDAP1}$  & $129 \pm 55$ &  '' & '' \\
$C$(FX$-$BX) & $135 \pm 5$ &  '' & '' \\
$E_\mathrm{act,QDAP}$ & $45 \pm 5$ & V$_\mathrm{Cu}$(A1) & 63 \\
FB$-$QDAP & $36 \pm 7$ &  '' & '' \\
$E_\mathrm{act,DDAP2}$ & $39 \pm 5$ & '' & '' \\
Eq.~\ref{eq:deep_defects} (DDAP1) & $300 \pm 50$ & A2 & ?? \\
Eq.~\ref{eq:deep_defects} (DDAP2) & $700 \pm 50$ & D2 & ?? \\
\hline
\end{tabular}
\caption{Defect analysis for CBTS. Defect ionization energies $E_\mathrm{i}$ are extracted from PL characterization. Charge transition levels $E_\mathrm{t}$ are calculated from first principles. The experimental ionization energies are derived either from thermal activation energies or from the shift between different peaks, as indicated in the table. Experimentally determined defect levels are tentatively assigned to theoretically predicted defects. $E_\mathrm{g}(T)$ is the estimated band gap at temperature $T$. The origin of the A2 and D2 defects is uncertain. Our interpretation is visualized in Fig.~\ref{fig:defect_levels}.}
\label{tab:defects_CBTS}
\end{table}

\subsubsection{Photoluminescence of CSTS}
As for the case of CBTS, interpretation of PL features in CSTS requires detailed analysis based on the data in Fig.~S2, Supporting Information. The results will be summarized in Table~\ref{tab:defects_CSTS}. Two main PL features are observed in CSTS at 83~K (Fig.~\ref{fig:T_dep}(b)): a feature at $\sim$2.0~eV with a clear high-energy shoulder, and a broad asymmetric feature in the 1.2-1.7~eV range. Each feature can be fitted with two Gaussian peaks. The positions of the Gaussian peaks in the 2.0~eV feature are about 1.96~eV and 2.06~eV, with negligible excitation intensity- and temperature dependence up to $\sim$150~K (Figs.~S2(a,e)). We assign the 1.96~eV peak to a TI transition (TI1) as in CBTS. Although it seems logical to identify the 2.06~eV as a FX or BB peak, there are two trends that are inconsistent with such an assignment. First, the TI1 peak has a higher $k$ coefficient than the 2.06~eV peak (Fig.~S2(d)), which is in contrast with the corresponding $k$ coefficients in CBTS (Fig.~S2(d)) and is not expected for a defect-related transition versus a non-defect-related transition.~\cite{Siebentritt2006} Second, the 2.06~eV peak remains at a constant position until 175~K and then it red-shifts by a much larger amount than the expected thermal band gap narrowing (Fig.~S1(a)). Hence, we assign the 2.06~eV peak to a band-to-tail (BT) transition instead. Similarly to the case of CBTS, the TI peak of CSTS blue-shifts in the 150-200~K temperature range and then follows the red shift of the FX peak above 200~K. As in CBTS, we interpret the $\sim$30~meV blue shift at intermediate temperatures as flattening of band edge fluctuations. However, there is an important qualitative difference between the two materials. Namely, the FWHM of most peaks in CSTS decreases with excitation intensity, instead of increasing as in CBTS. This behavior is compatible with electronic screening of band edge fluctuations by photocarriers, rather than state filling.~\cite{Rey2018,Lang2017} As band edge fluctuations flatten, the BT peak turns into a FX peak and the TI1 peak turns into a FB peak (FB1), which quenches and leaves the FX peak as the dominant peak at RT. The dominance of a FX peak in CSTS at RT is compatible with the excitonic feature observed in the absorption coefficient of CSTS at RT.~\cite{Crovetto2019a} As for CBTS, this calls for a re-evaluation of the band gap of CSTS using an Elliott function, since standard Tauc analysis is not applicable and results in band gap underestimation. The exciton binding energy of CSTS is estimated as 62~meV using our calculated static dielectric constant of 6.1 and previously calculated average effective masses (0.22~$m_0$ for electrons and 0.82~$m_0$ for holes).~\cite{Pandey2018,Zhu2017} Assuming the same temperature dependence of the band gap as in CBTS, the BT peak lies $\sim$24~meV below the estimated low-temperature band gap of CSTS, in good agreement with the tail state depth estimated by the blue shift of the TI peak ($\sim$30~meV). PL transitions involving defects that are shallower than the band tails would merge with tail emission, so the existence of shallow defects with ionization energy $<$30~meV cannot be confirmed nor excluded.

\begin{table}[t!]
\centering
\begin{tabular}{l l c c }
\hline
 \multicolumn{2}{l}{CSTS: Experiment} & \multicolumn{2}{l}{CSTS: Theory} \\
Exp. quantity  & $E_\mathrm{i}$ (meV) & Defect & $E_\mathrm{t}$ (meV)  \\
\hline
$E_\mathrm{g}$(83~K)$-$BT & $< 30$? & Cu$_\mathrm{i}$(D1?) & 0 \\
BT$-$TI1 & $108 \pm 4$ & V$_\mathrm{Cu}$(A1) & 70 \\
%$E_\mathrm{g}$(300~K)$-$FB1 & $122 \pm 15$ & " & ''  \\
FB2$-$DDAP & $99 \pm 9$ &  '' & ''  \\
$E_\mathrm{act,DDAP}$ & $108 \pm 41$ &  '' & ''  \\
BT$-$TI2 & $600 \pm 50$ & D2 & ?? \\
\hline
\end{tabular}
\caption{Defect analysis for CSTS, using the same symbols as in Table~\ref{tab:defects_CBTS}. Note that the D1 defect was not detected experimentally, but it might be concealed by band edge fluctuations due to its very low ionization energy. The origin of the D2 defect is uncertain. Our interpretation is visualized in Fig.~\ref{fig:defect_levels}}
\label{tab:defects_CSTS}
\end{table}

\begin{figure}[t!]
\centering%
\includegraphics[width=1\columnwidth]{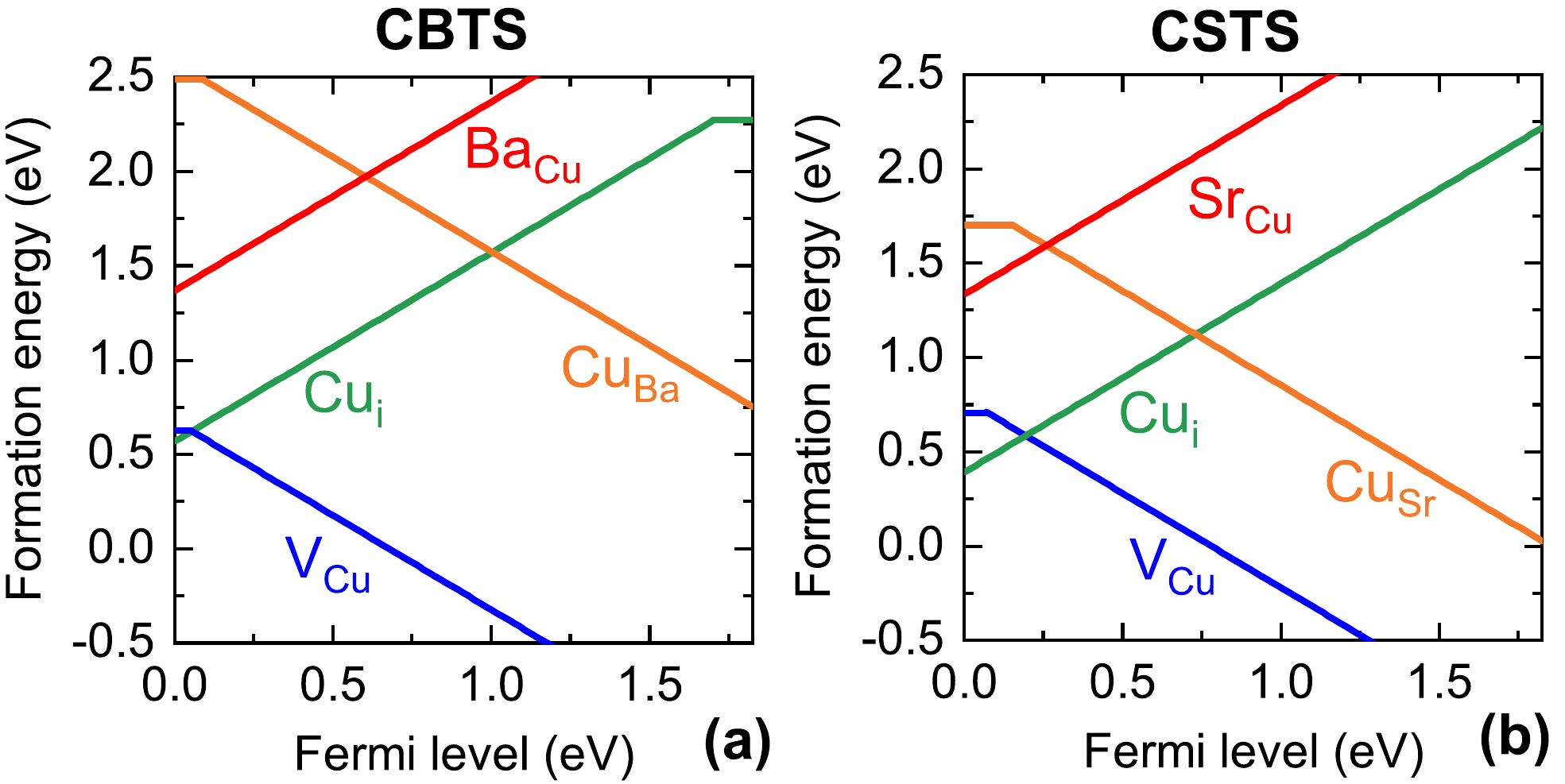}
\caption{Calculated formation energy of the V$_\mathrm{Cu}$, Cu$_\mathrm{i}$, Cu$_\mathrm{II}$, and II$_\mathrm{Cu}$ defects in CBTS and CSTS under II-rich conditions as a function of the Fermi level. A Fermi level of 0~eV corresponds to the valence band maximum. The calculated band gaps are 1.82~eV for both CBTS and CSTS.}
\label{fig:defects_fermi_level}
\end{figure}

The ionization energy of the (less shallow) defect involved in the TI1/FB1 transition can be estimated based on the shift between the BT and the TI1 peaks at low temperatures ($108 \pm 8$~meV) or from the activation energy of the FB1 peak. However, the latter has an extremely large standard error ($E_\mathrm{act} = 143 \pm 168$~meV) because the FB1 transition only begins to quench at relatively high temperatures (Fig.~S2(b)), so we discard it in our analysis.
The broad PL feature in the 1.2-1.7~eV range peak can also be fitted with two separate Gaussian peaks centered at 1.36~eV and 1.46~eV at 83~K. The peaks quench at different rates at higher temperatures, so that the lower-energy (high-energy) peak is dominant at low (high) temperatures. Above 160~K, the intensity of the low-energy peak is too low to be determined reliably, so we only fit the higher-energy peak with a single Gaussian function. Unlike the case of CBTS, this peak is still clearly visible at RT (Fig.~\ref{fig:T_dep}(b)). The 1.46~eV peak does not shift with excitation intensity or with increasing temperature, whereas a small blue shift with excitation intensity and temperature is observed for the 1.36~eV peak. Thus, we assign the 1.46~eV peak to a TI peak transition involving a deep impurity and we assign the 1.36~eV peak to a DDAP transition. Similarly to the TI1 peak, the TI2 peak turns into a FB transition (FB2) at higher temperatures where band edge fluctuations flatten out. Thermal quenching of the TI2/FB2 and DDAP peaks can be fitted with Eq.~\ref{eq:FB}, yielding activation energies of $69 \pm 24$~meV and $108 \pm 41$~meV for the TI2/FB2 peak and the DDAP peak, respectively (Fig.~S2(b)). The former activation energy cannot be related to ionization of the (much deeper) defect involved in the TI2/FB2 transition, and is probably related to the activation of a competing non-radiative recombination channel instead. The ionization energy of this defect can, however, be estimated as $600 \pm 25$~meV based on the shift between the BT peak and the TI2 peak. On the other hand, the activation energy of the DDAP peak ($108 \pm 41$~meV) is in good agreement with the ionization energy of the TI1/FB1 defect so we conclude that: (i) the shallow defect involved in the DDAP transition is the same defect responsible for the TI1/FB1 transition, and (ii) the deep defect involved in the DDAP transition is the same $\sim$600~meV defect responsible for the the TI2/FB2 transition. Consistent with this interpretation, the shift between the TI2 and the DDAP peak ($99 \pm 9$~meV) is in good agreement with the activation energy of the DDAP peak.

% ($122 /pm 8$~meV), with good agreement between the two methods. Due to the appreciable intensity decay of the TI/FB peak at low temperatures (Fig.~S2(b)), the activation energy was extracted based on the alternative equation~\cite{Krustok1997}
%\begin{equation}
%I_\mathrm{PL} = \frac{I_0}{1+aT^{3/2}\exp(-E_\mathrm{act}/k_\mathrm{B}T) + bT^{3/2}}
%\label{eq:Krustok}
%\end{equation}

To conclude, we have identified a shallow defect ($\sim$110~meV ionization energy) and a deeper defect of opposite type ($\sim$600~meV ionization energy) in CSTS. Since band edge fluctuations are present at low temperature, the existence of shallow defects with ionization energy $<$30~meV cannot be confirmed nor excluded. These results are summarized in Table~\ref{tab:defects_CSTS}. Similarly to CBTS, the tail states caused by band edge fluctuations are relatively shallow ($\sim$30~meV) and seem to disappear at temperatures higher than $\sim$200~K.

\begin{figure}[t!]
\centering%
\includegraphics[width=0.8\columnwidth]{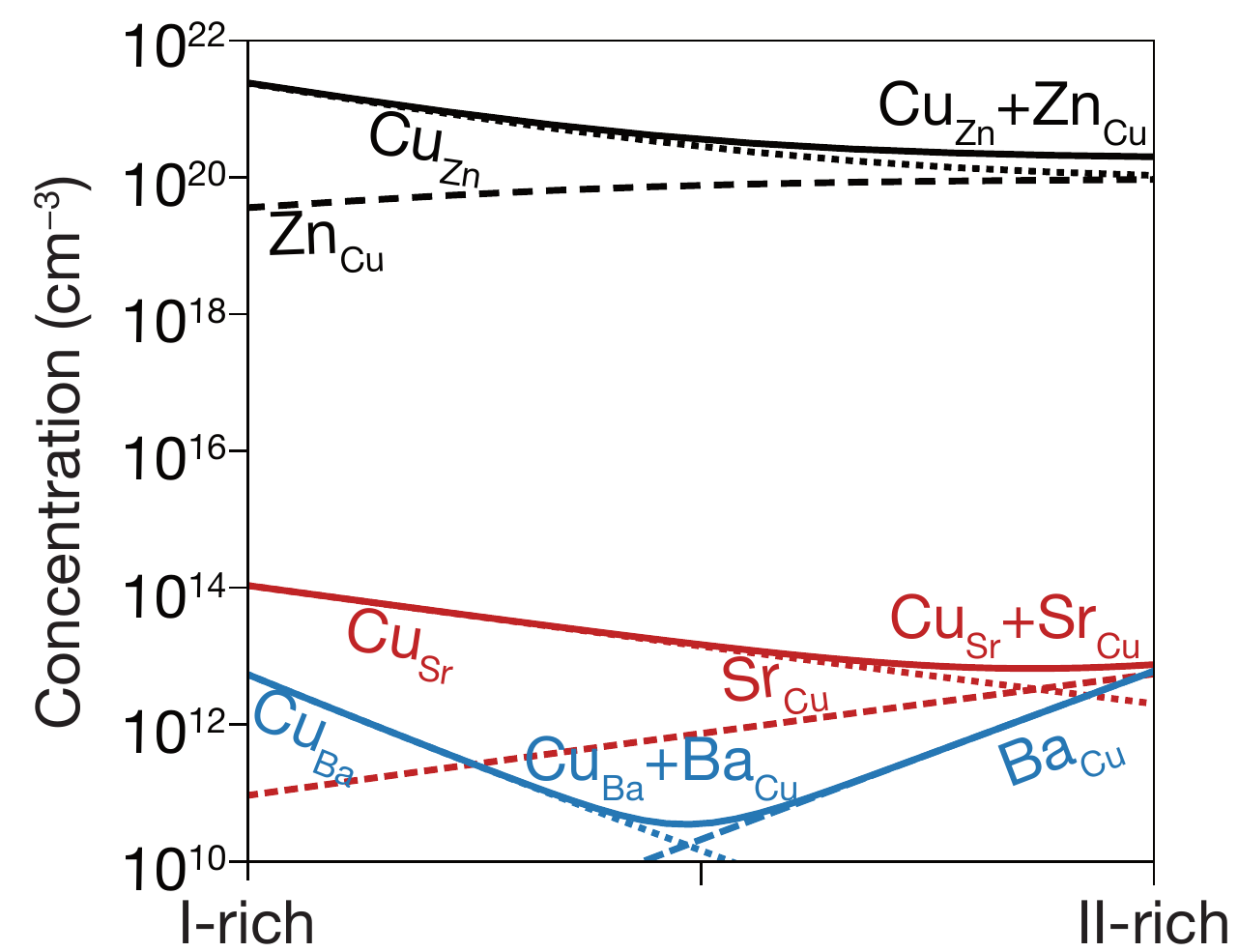}
\caption{Thermal equilibrium concentration of I$_\mathrm{II}$ and II$_\mathrm{I}$ antisite defects in CZTS (black lines), CSTS (red lines), and CBTS (blue lines) under I-rich and II-rich conditions. 
These are based on first-principles defect calculations.
The dashed and dotted lines represent concentrations of II$_\mathrm{I}$ and I$_\mathrm{II}$ antisites, respectively. Their sum is shown in solid lines.}
\label{fig:antisites}
\end{figure}

\subsection{Interpretation of defect levels in CBTS and CSTS}
%As mentioned in the previous sections, PL measurements can determine the ionization energy of the defects involved in radiative transitions, but it is difficult to distinguish between donors and acceptors. 
To investigate the possible chemical origin of the defects identified experimentally, we perform first-principles defect calculations. 
Previous work showed the single acceptor V$_\mathrm{Cu}$ and the single donor Cu$_\mathrm{i}$ to be the lowest-energy shallow defects in CBTS.~\cite{Hong2016}
We also find that V$_\mathrm{Cu}$ and Cu$_\mathrm{i}$ have shallow charge transition energies and much lower formation energies than the II$_\mathrm{Cu}$ donors and Cu$_\mathrm{II}$ acceptors (Fig.~\ref{fig:defects_fermi_level}), which are expected to be 7 orders of magnitude less abundant than in CZTS and CCTS (Fig.~\ref{fig:antisites}). 
The particularly low formation energy of V$_\mathrm{Cu}$ favors p-type conductivity in both materials, as observed by experiment.~\cite{Crovetto2019a,Ge2017} 
Compensation of V$_\mathrm{Cu}$ by the Cu$_\mathrm{i}$ donor is stronger in CSTS than in CBTS, which is consistent with a higher Fermi level position measured in CSTS with respect to its valence band maximum.~\cite{Crovetto2020b} 

\begin{figure}[t!]
\centering%
\includegraphics[width=\columnwidth]{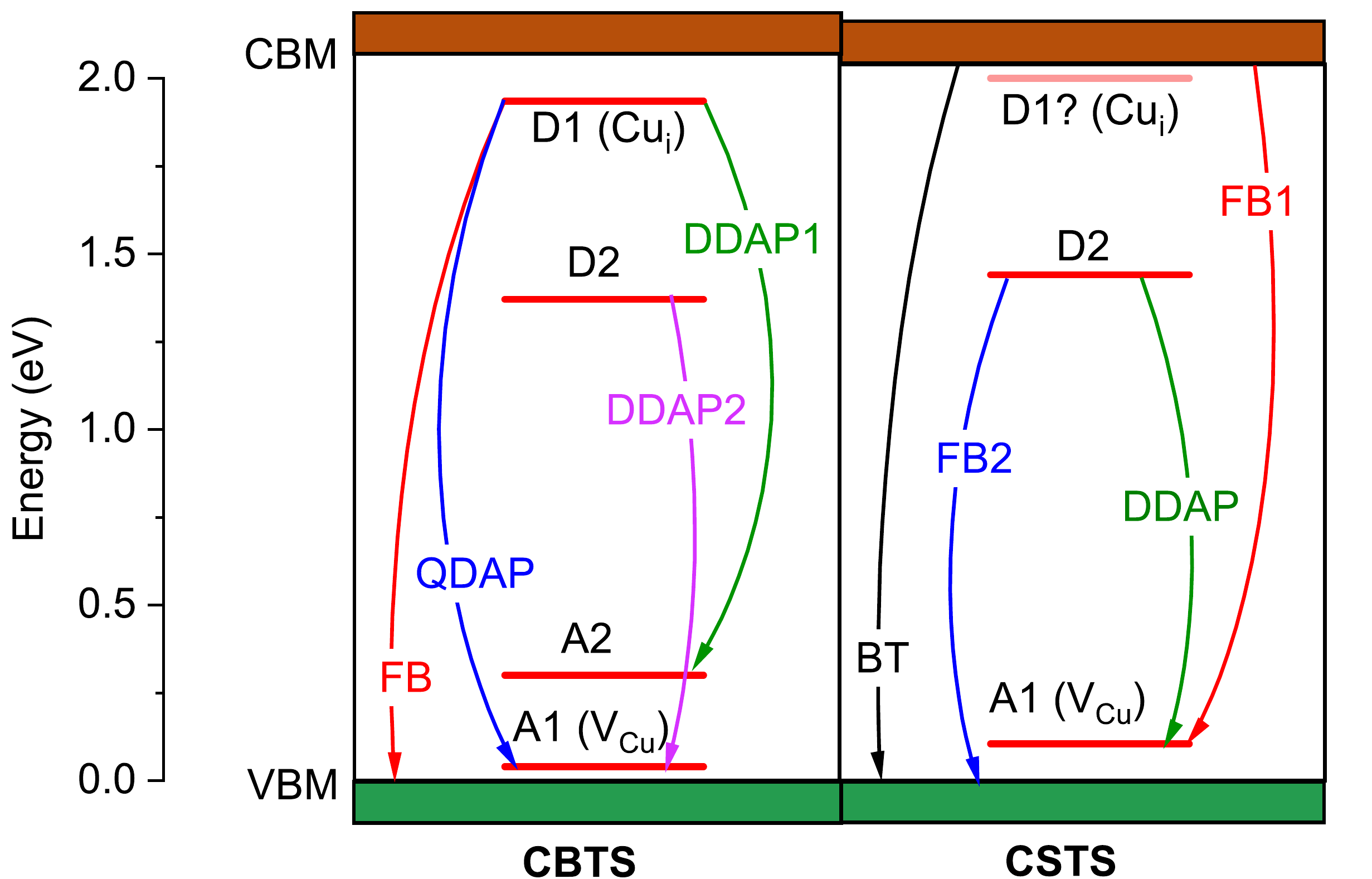}
\caption{Summary of the PL transitions identified in CBTS and CSTS. The defect levels inferred from such transitions are indicated as A$_\mathrm{x}$ if acceptors or D$_\mathrm{x}$ if donors. Identification of a given defect as a donor or an acceptor is done by analogy to the calculated charge transition levels of V$_\mathrm{Cu}$ and Cu$_\mathrm{i}$, as shown in Tables~\ref{tab:defects_CBTS},\ref{tab:defects_CSTS}.}
\label{fig:defect_levels}
\end{figure}

\begin{figure*}[t!]
\centering%
\includegraphics[width=\textwidth]{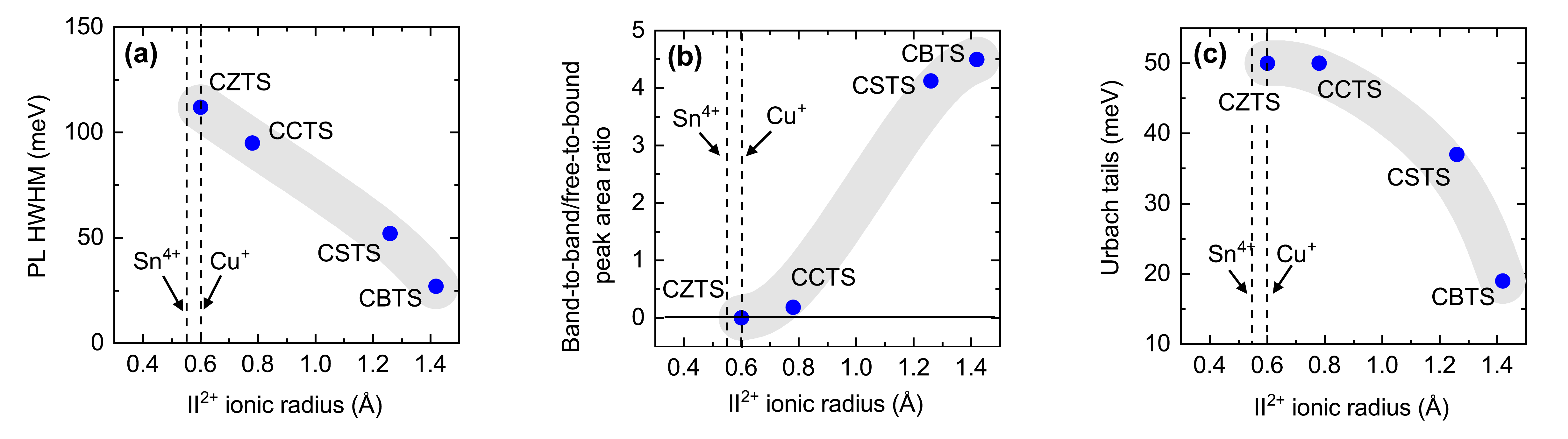}
\caption{Correlations between various band tail-related quantities and the Shannon radius of the II$^{2+}$ cation in CZTS, CCTS, CSTS, and CBTS. The radii of the Cu$^{+}$ and Sn$^{4+}$ cations common to all four compounds are shown for reference. The quantities taken as a measure of band tailing are: (a) the low-energy side half width at half maximum (HWHM) of the dominant PL peak at RT; (b) The integrated intensity ratio between the PL peaks identified as a free-to-bound transition and as a band-to-band (or exciton) transition; the Urbach energy extracted by EQE as shown in Fig.~\ref{fig:fluctuations_tails}(b). The shaded areas are a guide to the eye. All measurements were performed at room temperature.}
\label{fig:ionic_radius_vs_tails}
\end{figure*}

Since V$_\mathrm{Cu}$ and Cu$_\mathrm{i}$ are the most probable origin of the shallow defects identified by PL, we compare their calculated charge transition levels $E_\mathrm{t}$ to the measured defect ionization energies $E_\mathrm{i}$ in Table~\ref{tab:defects_CBTS} and Table~\ref{tab:defects_CSTS}. In CBTS, $E_\mathrm{t} = 63$~meV for V$_\mathrm{Cu}$ and $E_\mathrm{t} = 118$~meV for Cu$_\mathrm{i}$. These levels are in good agreement with the experimental ionization energies of $\sim$40~meV and $\sim$135~meV. Therefore, we assign the $\sim$40~meV level to the V$_\mathrm{Cu}$ acceptor and we assign the $\sim$130~meV level to the Cu$_\mathrm{i}$ donor as shown in Fig.~\ref{fig:defect_levels}. This assignment can be substantiated by analysis of the bound excition peak observed in CBTS (BX in Fig.~\ref{fig:T_dep}(a)). The bound exciton involves a defect with ionization energy $C$(BX-FX), where BX$-$FX = 27~meV is the shift between the two excitonic peaks in Fig.~\ref{fig:T_dep}(a) and $C$ is Haynes' constant, which depends on the defect type and can be determined from the ratio of the effective masses of CBTS.~\cite{Sharma1967,Atzmuller1979} Assuming the same effective masses that we used for the estimation of $E_\mathrm{b}$,~\cite{Zhu2017,Pandey2018} three possibilities exist. If the exciton was bound to an ionized impurity, the ionization energy of the latter would be $\sim$27~meV as $C \sim 1$ for both donors and acceptors.~\cite{Sharma1967} If the exciton was bound to a neutral impurity, the ionization energy of the latter would be $\sim$135~meV for a donor or $\sim$360~meV for an acceptor.~\cite{Atzmuller1979} The best match with the experimentally detected defects in Table~\ref{tab:defects_CBTS} is between the $\sim$135~meV neutral donor expected by Haynes's rule and the $\sim$135~meV defect found by PL analysis, which was also assigned to a donor based on our computational results. This donor is expected to be neutral at the temperature where the bound exciton peak is detected (79~K) because thermal quenching of the FB peak, corresponding to donor ionization, occurs at a much higher temperature.

The situation in CSTS is reversed, as the calculations predict V$_\mathrm{Cu}$ to be deeper than Cu$_\mathrm{i}$. The calculated transition energies of the two defects (70~meV and 0~meV) suggest that the experimentally-determined $\sim$105~meV defect may be attributed to the V$_\mathrm{Cu}$ acceptor. Note that Cu$_\mathrm{i}$ is predicted to be extremely shallow, so it is in practice invisible to PL characterization even if present in a large concentration because of band edge fluctuations in CSTS at low temperature.  Assignment of the deeper radiative levels to specific point defects is more difficult for both materials. The $\sim$300~meV acceptor in CBTS is potentially compatible with the charge transition levels of the the S$_\mathrm{i}$ and S$_\mathrm{Cu}$ single acceptors, which were identified as the lowest-energy deep acceptors in a previous calculation.~\cite{Hong2016} The ionization energies of the $\sim$700~meV and $\sim$600~meV deep donors found in CBTS and CSTS do not match the charge transition energy of any low-formation energy donors.~\cite{Hong2016} The origin of these defects is unknown.

\subsection{Defect trends in Cu$_2$-II-Sn-S$_4$ compounds}
The shallow defects identified from PL transitions of CZTS and CZTSe are often assigned to cation antisites, typically Zn$_\mathrm{Cu}$ or Cu$_\mathrm{Zn}$. Their formation energies and charge transition levels are similar to those of the equivalent Cd$_\mathrm{Cu}$ and Cu$_\mathrm{Cd}$ antisites in CCTS.~\cite{Yuan2015,Hadke2019} Furthermore, the characteristics of the room-temperature FB peak in CCTS are similar to those of the corresponding PL peak in CZTS~\cite{Hadke2019} so it likely that antisites are responsible for shallow defect PL emission in CCTS as well. On the other hand, the large size of the Sr$^{2+}$ and Ba$^{2+}$ cations combined with their unique coordination number in the $P3_1$ structure (Table~\ref{tab:ionic_radii}) makes formation of II$^{2+}$-based antisites energetically unfavorable (Fig.~\ref{fig:antisites}). Instead, V$_\mathrm{Cu}$ and Cu$_\mathrm{i}$ are likely to be the main shallow defects involved in radiative transitions in CBTS and CSTS, as discussed in the previous section.

%Both CBTS and CSTS are found to be native p-type semiconductors at RT.~\cite{Crovetto2019a,Crovetto2020b} Interestingly, the dominant acceptor in CBTS (V$_\mathrm{Cu}$) is shallower than the main compensating donor (Cu$_\mathrm{i}$). Conversely, the compensating donor in CSTS is shallower than the main acceptor (Fig.~\ref{fig:defect_levels}). Thus, CSTS may experience a transition from p-type to n-type conductivity at low temperatures, due to less effective ionization of the deeper acceptor. Finally, the shallow defects found by PL in CBTS and CSTS are shallower than the ones typically found in CZTS by analysis of activation energies and red-shifts with respect to the band gap.~\cite{Levcenko2016,Grossberg2019} This is consistent with the fact that deeper levels have been found for the Zn$_\mathrm{Cu}$ and Cu$_\mathrm{Zn}$ antisites in CTZS and CCTS compared to the  V$_\mathrm{Cu}$ and Cu$_\mathrm{i}$ levels in CBTS and CSTS. It must be remembered, however, that the large band gap/potential fluctuations in CZTS make it difficult to identify very shallow defects independently from the fluctuations themselves.

Apart from these differences in the dominant radiative defects, it is interesting to verify whether the ionic radius of the II$^{2+}$ cation in the Cu$_2$-II-Sn-S$_4$ series is a good descriptor of the extent of band tailing in these materials, as originally hypothesized.~\cite{Hong2016} We start by observing that lower band tailing in CBTS and CSTS with respect to the rest of the series cannot simply be explained by better dielectric screening, because their static dielectric constant is lower than in CZTS (Table~\ref{tab:ionic_radii}). Three correlations can instead be provided, which seem to confirm the hypothesized role of the cationic radius. The first (Fig.~\ref{fig:ionic_radius_vs_tails}(a)) is a simple correlation between the low-energy-side half-width at half-maximum (HWHM) of the main room-temperature PL peak and the II$^{2+}$ ionic radius. The low-energy side of the PL spectrum corresponds to the density of the states involved in PL emission. In a tail-free material, the PL HWHM is simply $k_\mathrm{B}T/2$ due to the thermal distribution of carriers in the bands.~\cite{Siebentritt2006} In the presence of tail states, the HWHM increases to reflect the tail density of states. The behavior of the PL HWHM across the Cu$_2$-II-Sn-S$_4$ series confirms that tail states become shallower as the ionic radius of the II$^{2+}$ cation deviates more and more from the ionic radii of Cu$^{+}$ and Sn$^{4+}$. A second correlation (Fig.~\ref{fig:ionic_radius_vs_tails}(b)) involves the relative weight of the FB and the BB (or exciton) PL peaks at RT. In the presence of tail states or shallow defects that recombine radiatively, the intensity of the BB peak is reduced due to competition with tail- and defect-related transitions. Again, the trend in the Cu$_2$-II-Sn-S$_4$ series is consistent with a decrease in tail state- and shallow defect density as the I$^{+}$-II$^{2+}$ cation size mismatch increases going from Zn$^{2+}$ to Ba$^{2+}$. Finally, Urbach and Gaussian tails of the various compounds can be extracted directly from analysis of the solar cell's external quantum efficiency (EQE) as shown in Fig.~\ref{fig:fluctuations_tails}. Plotting the Urbach energy versus the II$^{2+}$ ionic radius results again in the expected trend. Note that the Gaussian band gap distribution is also a decreasing function of the ionic radius, although Gaussian tails of CZTS are much larger, for reasons that will be discussed later. In conclusion, the tail state chemical trend in the Cu$_2$-II-Sn-S$_4$ series suggests that introducing size-mismatched cations is a successful strategy to mitigate band tailing in Cu$_2$-II-Sn-S$_4$ compounds, which further indicates that some type of antisite defect is the main cause of band tails in this class of materials. In the next section, the specific origin of band tails is investigated.

\subsection{Tail state trends in kesterite-inspired materials}
We now extend our discussion of band tails to include other kesterite-inspired materials involving substitution of Cu, Sn, and S, and attempt to understand the origin of band tails based on chemical trends. The origin of band tails in CZTS and CZTSe has been a heavily debated subject, which has not been entirely resolved. Two basic issues are important. The first is whether to attribute band tails to potential fluctuations or to band gap fluctuations. Potential fluctuations are the classical mechanism behind band tail formation in heavily compensated semiconductors. They are caused by a high concentration and non-random distribution of acceptors and donors, causing fluctuations in the electrostatic potential and non-local PL transitions between spatially separated potential wells for electrons and holes.~\cite{Shklovskii1984} Band gap fluctuations could be caused by the precipitation of binaries or ternary phases,~\cite{Siebentritt2012} by competition between different crystal structures with similar energy and different band gaps (e.g. kesterite and stannite),~\cite{Persson2010,Siebentritt2012} or by the band gap narrowing effect of certain defect clusters.~\cite{Chen2013} The most recent studies seem to suggest that band gap fluctuations, rather than potential fluctuations, are responsible for the majority of band tails in CZTSe.~\cite{Rey2018,Lang2017} These conclusions are mainly based on observations of the broadening of the low-energy side of the PL spectrum versus excitation intensity. If potential fluctuations were the only contribution to band tails, increased electrostatic screening due to the increasing free carrier concentration would cause peak narrowing with increasing excitation intensity. If band gap fluctuations were the only contribution, increased state filling with increasing excitation intensity would cause either no change to the low-energy side width, or possibly some broadening with increasing excitation intensity due to contributions from a more diverse set of defects.~\cite{Rey2018} However, no or very limited narrowing is observed experimentally, pointing to the predominant role of band gap fluctuations.~\cite{Rey2018,Lang2017}

\begin{figure}[t!]
\centering%
\includegraphics[width=\columnwidth]{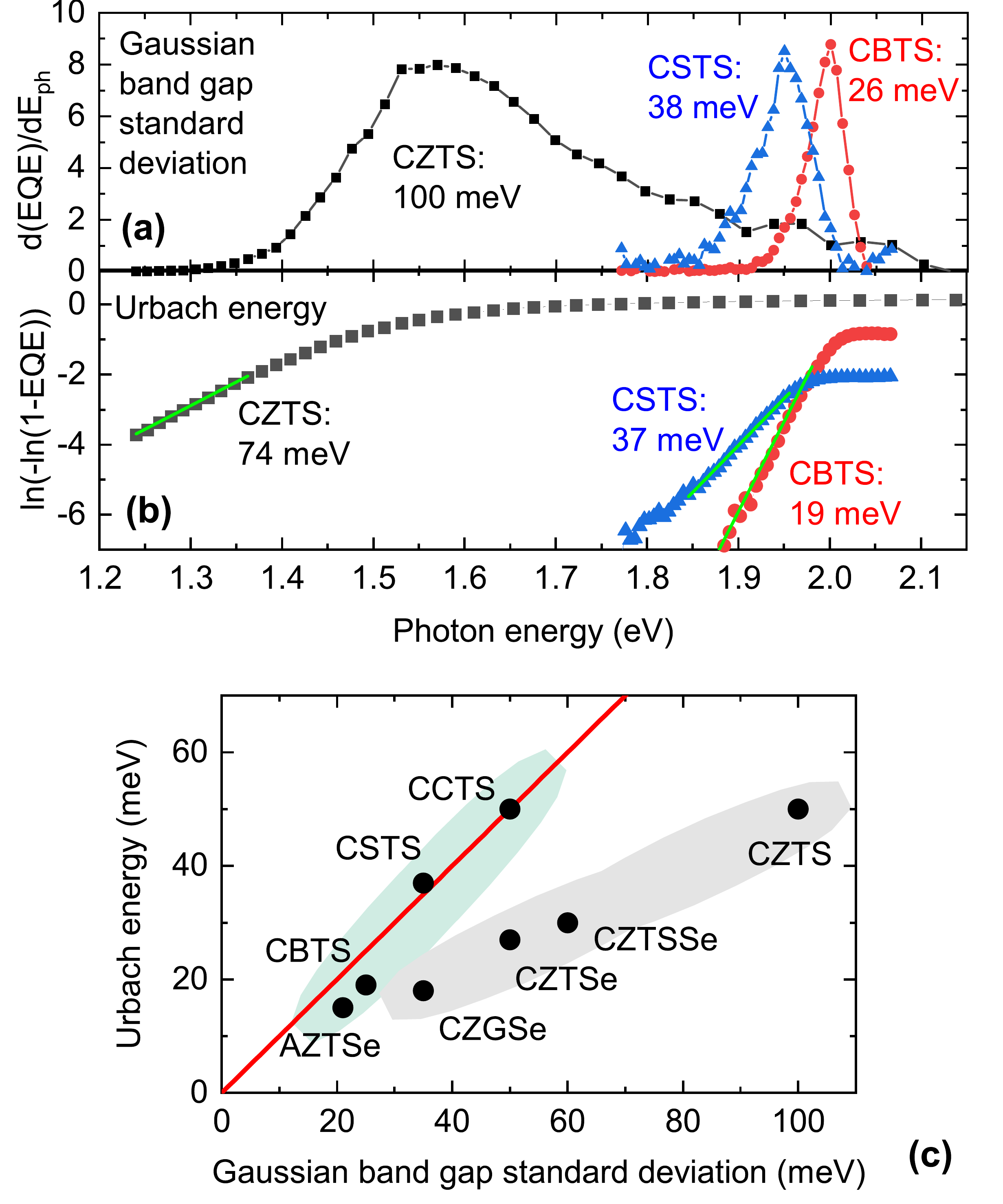}
\caption{(a) Extraction of the Gaussian band gap standard deviation for in-house-fabricated CZTS, CSTS and CBTS by external quantum efficiency (EQE) analysis. 
(b) Extraction of the Urbach energy for the same samples by EQE analysis. 
(c) Plot of the Urbach energy and Gaussian band gap standard deviation for a range of kesterite-inspired compounds. The data is compiled from various publications.~\cite{Crovetto2019a,Crovetto2019b,Gershon2016,Sahayaraj2017,Hadke2019,Rey2018}
Along the red line, the Urbach energy is approximately equal to the Gaussian band gap standard deviation. 
The two shaded areas are guides to the eye for grouping the materials with roughly similar Gaussian and Urbach tails, versus the materials in which the band gap standard deviation is larger than the Urbach energy.}
\label{fig:fluctuations_tails}
\end{figure}

The second issue in the band tail discussion is its chemical origin. Some researchers~\cite{Gershon2013,Scragg2016,Nishiwaki2018} have attributed band tailing to Cu-Zn disorder owing to the abundance of the (Cu$_\mathrm{Zn}$+Zn$_\mathrm{Cu}$) defect cluster, evidenced in Fig.~\ref{fig:antisites}. This can result both in band gap fluctuations due to the narrowing effect of the (Cu$_\mathrm{Zn}$+Zn$_\mathrm{Cu}$) cluster on the band gap,~\cite{Scragg2016} or in potential fluctuations due to the non-random distribution of Cu$_\mathrm{Zn}$ acceptors and Zn$_\mathrm{Cu}$ donors.~\cite{Gershon2013,Gokmen2013}
However, other researchers have shown that band tails are independent of the order parameter $S$ of kesterite, thus excluding a major involvement of the (Cu$_\mathrm{Zn}$+Zn$_\mathrm{Cu}$) cluster.~\cite{Bourdais2016,Rey2018} An often-invoked alternative chemical origin of the band tails is the (2Cu$_\mathrm{Zn}$+Sn$_\mathrm{Zn}$) defect cluster.~\cite{Rey2018,Hadke2019,Ma2019} According to first-principles calculations,~\cite{Chen2013} this defect cluster has a low formation energy and causes significant band gap narrowing in CZTS. Its role in the experimentally observed band tails in kesterite is plausible but has not been proven experimentally so far.

In this section, we will take a different approach to the analysis of the causes of band tailing in kesterite-related materials. Our method is based on the analysis of band tail data over a chemical space containing various kesterite-inspired materials, and its comparison to calculated defect properties for those materials. 
This approach relies on the recent publication of experimental band tail data for various fully-substituted, kesterite-inspired semiconductors including CZTS,~\cite{Hadke2019,Yan2017a,Nishiwaki2018} CZTSe,~\cite{Rey2018,Hages2016,Nishiwaki2018} CZTSSe,~\cite{Rey2018,Gershon2016,Gokmen2013} CCTS,~\cite{Hadke2019} Ag$_2$ZnSnSe$_4$ (AZTSe),~\cite{Gershon2016} and Cu$_2$ZnGeSe$_4$ (CZGSe).~\cite{Sahayaraj2017,Nagaya2018} 
It also relies on the availability of detailed defect calculations for CZTS, CZTSe, CCTS, and Ag$_2$ZnSnS$_4$ (AZTS).~\cite{Chen2013,Yuan2015}
The results of these calculations can be compared quantitatively, since they were performed by the same group using the same computational approach.  Formation energies and band gaps of competing crystal structures are also available for various kesterite-inspired materials as a single consistent data set.~\cite{Chen2010c,Pandey2018}

Two clearly distinct types of experimental band tail data can be found. The first is a measure of the abruptness of optical absorption around the absorption onset, which can be derived either by optical measurements or by external quantum efficiency (EQE) measurements. To obtain an explicit quantity, one can take the derivative of EQE with respect to photon energy from below the absorption onset and up to its first inflection point (i.e., the maximum of the derivative) and fit it with a half-Gaussian function. The standard deviation of the Gaussian function can be interpreted as the standard deviation of the band gap distribution in the material.~\cite{Mattheis2007} In Fig.~\ref{fig:fluctuations_tails}(a) we perform this analysis on the EQE of CBTS, CSTS, and CZTS solar cells presented in previous work.~\cite{Crovetto2019a,Crovetto2019b,Cazzaniga2017} Clearly, the Gaussian distribution of band gaps in CSTS and especially in CBTS is much narrower than in CZTS.

The second type of band tail data involves plotting $\ln[-\ln(1-\mathrm{EQE})]$ at photon energies below the main absorption onset and fitting it with a straight line to extract the Urbach energy, which is the characteristic constant of a single-exponential tail.~\cite{Hages2016} A similar method can be employed using the absorption coefficient of the material on a logarithmic scale, although a measurement technique which is sensitive to very low absorption coefficients must be used.~\cite{Rey2018a,Malerba2014} In Fig.~\ref{fig:fluctuations_tails}(b) we perform this analysis on the EQE measured on the same CBTS, CSTS, and CZTS solar cells as above. The result is qualitatively similar to the Gaussian analysis, with CBTS having a lower Urbach energy than CSTS, which in turn has a much lower Urbach energy than CZTS.

\begin{figure*}[t!]
\centering%
\includegraphics[width=\textwidth]{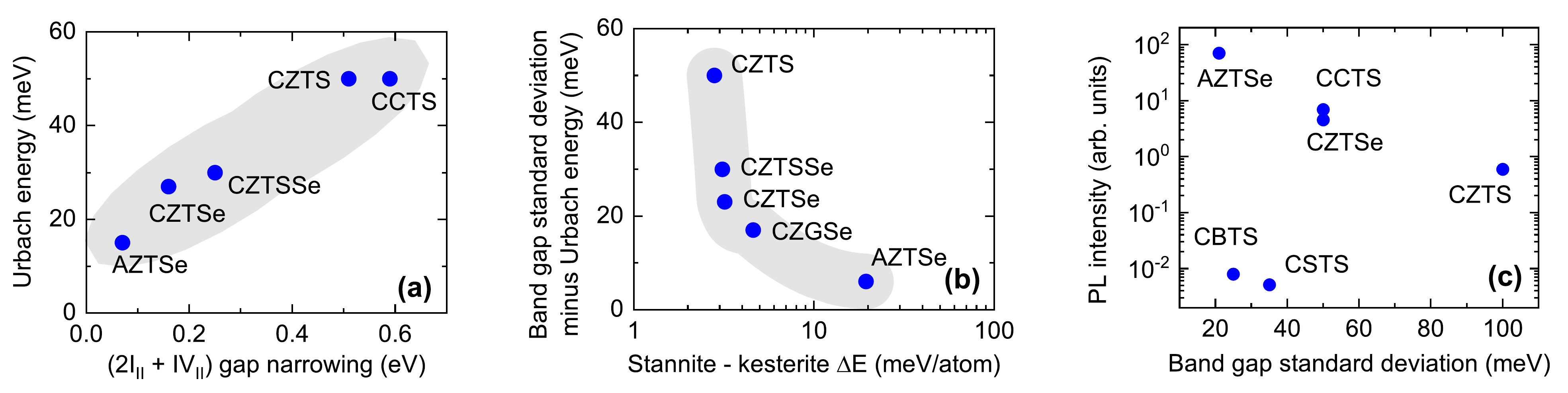}
\caption{(a): Correlation between the experimentally measured Urbach energy and the computationally determined band gap narrowing caused by one (2I$_{\mathrm{II}}$ + IV$_{\mathrm{II}}$) defect cluster per 128 atoms. Only for the case of AZTSe, the (2I$_{\mathrm{II}}$ + IV$_{\mathrm{II}}$) defect cluster has a very high formation energy so the (IV$_{\mathrm{II}}$ + II$_{\mathrm{IV}}$) cluster is considered instead since it causes the largest band gap narrowing among low-formation-energy clusters.~\cite{Yuan2015} Further details are available in the Supporting Information. (b): Correlation between the Gaussian band gap standard deviation in excess of the Urbach energy and the calculated energy difference between the stannite and kesterite polymorphs ($\Delta E_{\mathrm{f}}$ in Table~\ref{tab:competing_structures}). (c) Plot of the integrated PL intensity of kesterite-inspired materials versus their band gap standard deviation, showing that non-radiative losses and band tailing are uncorrelated. Non-radiative losses decrease by 59~meV per decade of PL intensity. In all cases, shaded areas are guides to the eye.}
\label{fig:correlations}
\end{figure*}

The Urbach energies and Gaussian band gap standard deviations measured for various kesterite-inspired materials are plotted against each other in Fig.~\ref{fig:fluctuations_tails}(c). Materials based on substitutions of all four elements are included. We observe that the Urbach energy and the band gap standard deviation coincide only for certain materials, i.e., the ones that line up on the red line in Fig.~\ref{fig:fluctuations_tails}(c). For other materials (CZGSe, CZTSe, and CZTS) the band gap standard deviation is larger than the Urbach energy by various amounts. Based on this trend, we propose that the Gaussian band gap distribution and the Urbach tails in CZTS, CZTSe, and CZGSe are not simply two possible models to quantify the same tailing phenomenon.~\cite{Gokmen2013} Instead, Gaussian tails and Urbach tails are two separate phenomena which have distinct physical origins and coexist in (as a minimum) CZGSe, CZTSe and CZTS. As will be shown in the following paragraphs, establishing this distinction is very important in order to find the most plausible causes of band tails in kesterite-inspired materials. Our approach to identify the origin of both types of tails is to correlate either the band gap standard deviation or the Urbach energy to computationally available quantities related to defects or competing phases.  Among them, we considered the following: (1) the calculated formation energy of various point defects and defect clusters that are known for CZTS, CZTSe, CCTS, and AZTS, such as I$_{\mathrm{II}}^-$, II$_{\mathrm{I}}^+$, IV$_{\mathrm{II}}$, (I$_{\mathrm{II}}$ + II$_{\mathrm{I}}$), (2I$_{\mathrm{II}}$ + IV$_{\mathrm{II}}$);~\cite{Chen2013,Yuan2015} (2) the calculated band gap narrowing (one defect per 128 crystal atoms) of various defect clusters, such as the (I$_{\mathrm{II}}$ + II$_{\mathrm{I}}$) and (2I$_{\mathrm{II}}$ + IV$_{\mathrm{II}}$) listed above;~\cite{Chen2013,Yuan2015} (3) the energy difference between the lowest-energy crystal structure and the second-lowest energy crystal structure;~\cite{Chen2010c,Pandey2018} and (4) the difference between the band gaps of the lowest-energy crystal structure and of the second-lowest energy crystal structure.~\cite{Chen2010c,Pandey2018}

We find that Urbach tails in CZTS, CZTSe, CCTS, and AZTSe are clearly not correlated to any of the investigated quantities, with the exception of band gap narrowing caused by the (2I$_{\mathrm{II}}$ + IV$_{\mathrm{II}}$) defect cluster (Fig.~\ref{fig:correlations}(a)). This correlation can be rationalized by remembering that Urbach tails are usually appropriate to model tail states that extend relatively deep into the forbidden gap.~\cite{Rey2018} In comparison to the (I$_{\mathrm{II}}$ + II$_{\mathrm{I}}$) cluster, the (2I$_{\mathrm{II}}$ + IV$_{\mathrm{II}}$) cluster is less abundant but has a much larger influence on the local band gap.~\cite{Chen2013} The observed correlation is in line with the assignment of several authors, e.g. Rey et al,~\cite{Rey2018} and Hadke et al.,~\cite{Hadke2019} who proposed (2Cu$_{\mathrm{Zn}}$ + Sn$_{\mathrm{Zn}}$) as the main responsible for band tailing in kesterite. The large difference between the calculated gap narrowing (hundreds of meV) and the measured Urbach energies (tens of meV) need not be regarded as an inconsistency, because band gap narrowing was calculated assuming a certain defect concentration (one cluster per 128 crystal atoms) which is higher than the expected concentration of the (2I$_{\mathrm{II}}$ + IV$_{\mathrm{II}}$) cluster given its formation energy. Furthermore, Urbach tails describe exponentially decaying states, so the Urbach energy does not represent the deepest states involved in band tailing.

Since Gaussian tails are known to give a better description of relatively shallow tails with a large density of states,~\cite{Rey2018} one may expect to find some sort of correlation between the band gap standard deviation and the properties of the (I$_{\mathrm{II}}$ + II$_{\mathrm{I}}$) cluster or of the corresponding individual defects. However no such correlation is found, with the main inconsistency being the behavior of CCTS. More precisely, the (I$_{\mathrm{II}}$ + II$_{\mathrm{I}}$) cluster has about the same formation energy and gap narrowing effect in CCTS and in CZTS,~\cite{Yuan2015,Chen2013,Hadke2019} but the band gap standard deviation in CCTS is about one half of the one in CZTS. On the other hand, we find a correlation between the standard deviation of band gaps and the energy difference between the ground state structure and the second lowest-energy structure (Fig.~\ref{fig:correlations}(b)), but only for the materials in which the latter has a lower band gap than the former. In all the materials that meet this requirement (CTZS, CZTSe, AZTSe, CZGSe) this corresponds to the energy difference between the ground state kesterite structure ($I\bar{4}$) and the stannite structure ($I\bar{4}2m$), as shown in Table~\ref{tab:competing_structures}. Conversely, in CCTS the ground state is stannite and the second lowest-energy structure is kesterite. In CBTS and CSTS, the ground state is $P3_1$ and the second lowest-energy structure is $P1n1$, which has a wider band gap (Table~\ref{tab:competing_structures}).
This correlation can be rationalized as follows. All the materials that line up on the red line in Fig.~\ref{fig:fluctuations_tails}(c) are the materials in which the most stable structure has a lower band gap than the second most stable structure (CSTS, CBTS), or in which the second most stable structure has a large energy difference with the most stable structure (AZTSe). In these materials, inclusion of the second most stable structure is either negligible in concentration, or it only modifies the absorption coefficient of the single-phase material above its absorption onset and therefore band tails are not detected using the method shown in Fig.~\ref{fig:fluctuations_tails}(a). Importantly, inclusion of the second structure does not result in potential wells that can trap carriers, such as the previously described (2I$_{\mathrm{II}}$ + IV$_{\mathrm{II}}$) defect cluster.
For the materials that deviate from the red line in Fig.~\ref{fig:fluctuations_tails}(c), the ease of formation of the lower-band gap stannite polymorph is a good descriptor of the extra Gaussian contribution to the band tails as illustrated in Fig.~\ref{fig:correlations}(b).

This correlation requires further discussion. As mentioned above, the (Cu$_\mathrm{Zn}$+Zn$_\mathrm{Cu}$) defect cluster has often been blamed for band tails in kesterite,~\cite{Gershon2013,Scragg2016,Nishiwaki2018} a hypothesis that has been disproved by several authors.~\cite{Rey2018,Malerba2017} However, closer examination reveals that many possible configurations of (Cu$_\mathrm{Zn}$+Zn$_\mathrm{Cu}$) clusters can exist, depending on the location of the swapping atoms.~\cite{Scragg2016,Nishiwaki2018} The lowest-energy configuration involves swapping of Cu and Zn atoms within the Cu-Zn planes (1/4 or 3/4) of the kesterite structure (Fig.~\ref{fig:structures}(a)\red{)}. We will label this \textit{in-plane} cluster configuration as (Cu$_\mathrm{Zn}$+Zn$_\mathrm{Cu}$)$_{\parallel}$. The (Cu$_\mathrm{Zn}$+Zn$_\mathrm{Cu}$)$_{\parallel}$ cluster is the one reported in Fig.~\ref{fig:antisites} of the present paper and in our reference defect calculations.~\cite{Yuan2015,Chen2013} The (Cu$_\mathrm{Zn}$+Zn$_\mathrm{Cu}$)$_{\parallel}$ configuration is responsible for the much-discussed cation disorder in CZTS, as it can be sufficiently abundant to be detected directly by, e.g., neutron diffraction.~\cite{Schorr2019} In fact, complete cation disorder in the 1/4 and 3/4 planes is achieved when there is one (Cu$_\mathrm{Zn}$+Zn$_\mathrm{Cu}$)$_{\parallel}$ cluster per 16 atoms (or two unit cells). Among all the conceivable (Cu$_\mathrm{Zn}$+Zn$_\mathrm{Cu}$) cluster configurations and concentrations, this completely disordered configuration has the lowest energy, only 0.3~meV/atom higher than the defect-free CZTS kesterite structure.~\cite{Scragg2016} However, many other cluster configurations and concentrations are possible.
A particularly interesting configuration is the \textit{out-of-plane} (Cu$_\mathrm{Zn}$+Zn$_\mathrm{Cu}$)$_{\perp}$ cluster, where one Cu atom from a Cu-Sn plane (0, 1/2, or 1) swaps with a Zn atom from a Cu-Zn plane. Formation of one (Cu$_\mathrm{Zn}$+Zn$_\mathrm{Cu}$)$_{\perp}$ cluster per 64 atoms (8 unit cells) costs 3.7~meV/atom more than forming a defect-free CZTS kesterite structure.~\cite{Scragg2016} There are two important features of the (Cu$_\mathrm{Zn}$+Zn$_\mathrm{Cu}$)$_{\perp}$ cluster. First, its formation energy is low but not as low as for the (Cu$_\mathrm{Zn}$+Zn$_\mathrm{Cu}$)$_{\parallel}$ cluster, indicating that it should be present in very high concentrations with respect to most other defects but it may be still difficult to detect in the refinement step of x-ray or neutron diffraction experiments. Second, the presence of (Cu$_\mathrm{Zn}$+Zn$_\mathrm{Cu}$)$_{\perp}$ clusters can be interpreted as a partial transition from the kesterite to the stannite structure. In fact, it can be seen from Fig.~\ref{fig:structures} that the stannite structure is equal to a kesterite structure with one (Cu$_\mathrm{Zn}$+Zn$_\mathrm{Cu}$)$_{\perp}$ defect cluster per unit cell.

From the above discussion, we propose that the additional Gaussian band tails observed in materials with the kesterite crystal structure are due either to "true" stannite inclusions or to a high density of (I$_\mathrm{II}$+II$_\mathrm{I}$)$_{\perp}$ defect clusters, which are related to a partial transition to a stannite structure. The formation energy of (I$_\mathrm{II}$+II$_\mathrm{I}$)$_{\perp}$ clusters is only available for CZTS so it cannot be used for correlation purposes across multiple materials. However, the energy difference between stannite and kesterite structures can be considered a good descriptor of such a formation energy and can support the role of (I$_\mathrm{II}$+II$_\mathrm{I}$)$_{\perp}$ clusters in the correlation observed in Fig.~\ref{fig:correlations}(b). These hypotheses are compatible with the outcome of previous studies~\cite{Rey2018,Malerba2017} which concluded that band tails in CZTS and CZTSe do not depend on the order parameter. The reason is that the order parameter is sensitive to (Cu$_\mathrm{Zn}$+Zn$_\mathrm{Cu}$)$_{\parallel}$ cluster concentrations but not necessarily to (Cu$_\mathrm{Zn}$+Zn$_\mathrm{Cu}$)$_{\perp}$ cluster concentrations or to the fraction of stannite present in the kesterite matrix. However, in a more general sense the I$_\mathrm{II}$ and II$_\mathrm{I}$ defects are still responsible for the Gaussian contribution to the band tails in the materials with a kesterite ground state structure CZTS, CZTSe, CZGSe, and AZTSe. Since the ground state structure of CCTS is stannite, Gaussian tails in CCTS are much less pronounced than in CZTS despite the nearly identical formation energy of the (I$_\mathrm{II}$+II$_\mathrm{I}$)$_{\parallel}$ cluster in the two materials.

\begin{table}[t!]
\centering
\begin{tabular}{l c c c }
\hline
%& \multicolumn{2}{c}{Ref.~\cite{Chen2010c}} & \multicolumn{2}{c}{Ref.~\cite{Pandey2018}} \\
& Ref.~\citenum{Chen2010c} & Ref.~\citenum{Pandey2018} & Ref.~\citenum{Pandey2018}  \\
& $\Delta E_{\mathrm{f}}$ & $\Delta E_{\mathrm{f}}$ &  $\Delta E_{\mathrm{g}}$  \\
material & (meV/at.) & (meV/at.) &  (meV/at.)  \\
\hline
AZTSe & $19.5$ & $18.1$  & $-230$    \\
CZGSe & $4.6$ & $5.4$ & $-240$    \\
CZTSe & $3.3$ & $3.7$ & $-170$    \\
CZTS & $2.8$ &  $2.9$ & $-150$   \\
\textbf{CCTS} & $\textbf{3.2}$ & $\textbf{4.7}$  & $\textbf{+100}$   \\
\textbf{CSTS} &  & $\textbf{35.2}$  & $\textbf{+100}$    \\
\textbf{CBTS} & & $\textbf{51.1}$  & $\textbf{+130}$    \\
\hline
\end{tabular}
\caption{Calculated energy difference $\Delta E_{\mathrm{f}}$ and band gap difference $\Delta E_{\mathrm{g}}$ between the two lowest-energy crystal structures of various kesterite-inspired materials. When the band gap of the lowest-energy structure is wider than the band gap of the second lowest-energy structure, $\Delta E_{\mathrm{g}}$ is taken with a negative sign. Otherwise, $\Delta E_{\mathrm{g}}$ is taken as positive and the material is labeled in bold font. Note that in all materials with a negative $\Delta E_{\mathrm{g}}$, the lowest-energy structure is kesterite and the second lowest-energy structure is stannite.}
\label{tab:competing_structures}
\end{table}

\subsection{Deep defects and non-radiative recombination trends in kesterite-inspired materials}
Fig.~\ref{fig:correlations}(c) shows the relative PL intensity of various kesterite-inspired semiconductors. According to Planck's law,~\cite{Wurfel2010} PL intensity depends exponentially on the splitting $\Delta \mu$ between the electron and hole quasi-Fermi levels in the illuminated semiconductor, following the approximate relation $\Delta \mu = k_\mathrm{b}T \ln(a\, I_\mathrm{PL})$, or $\ln(I_\mathrm{PL}) = \Delta \mu / k_\mathrm{b}T  - \ln(a)$. Here, $k_\mathrm{b}T$ is the thermal energy and $a$ is a constant reflecting the fact that the the PL intensities reported in Fig.~\ref{fig:correlations}(c) are in arbitrary units instead of as a quantum yield. Defining the voltage loss due to non-radiative recombination as $\Delta V_{\mathrm{nonrad}} =  V_\mathrm{oc}^\mathrm{SQ} - \Delta \mu$, where $V_\mathrm{oc}^\mathrm{SQ}$ is the maximum open-circuit voltage according to the Shockley-Queisser limit, it follows that every order of magnitude increase in $I_\mathrm{PL}$ corresponds to a $k_\mathrm{b}T \ln(10) \simeq 59$~mV decrease in the non-radiative loss $\Delta V_{\mathrm{nonrad}}$. Comparing PL intensities across various materials is then equivalent to comparing differences in their non-radiative voltage losses.

Note that the data in Fig.~\ref{fig:correlations}(c) was compiled from different sources, where the PL intensities of various absorbers were measured relative to one another.~\cite{Hadke2019,Gurieva2020,Crovetto2019a,Crovetto2019b} Thus, the data is to be intended only as semi-quantitative, as error bars up to one order of magnitude are expected for the relative PL intensity of material pairs that were not compared directly in the same study. It should also be noted that different excitation intensities may have been used in the different studies. However, the $k$ exponents in the $I_\mathrm{PL} \propto I^k_\mathrm{ex}$ power laws at RT are similar (around 1.3-1.4) for most of the materials shown in Fig.~\ref{fig:correlations}(c). This ensures that the relative intensity trends are roughly preserved using different excitation intensities.

Some interesting qualitative conclusions can be drawn from Fig.~\ref{fig:correlations}(c). First, PL intensity is clearly not correlated to the measured band gap standard deviation, indicating that non-radiative recombination and band tails are independent of each other. In other words, a given material may have a low density of defects responsible for band tailing but that doesn't imply a low density of efficient non-radiative recombination centers. For example, band tails in CBTS are quite shallow compared to most other kesterite-inspired materials but its non-radiative recombination loss $\Delta V_{\mathrm{nonrad}}$ is about 115~mV larger than in CZTS since its PL intensity is about two orders of magnitude lower. A second important conclusion is related to the materials design principle based on intentionally employing size-mismatched cations to discourage the formation of antisite defects. 
The introduction of the large Ba$^{2+}$ and Sr$^{2+}$ cations in CBTS and CSTS does prevent the formation of some antisite defects (Figs.~\ref{fig:antisites} and  \ref{fig:ionic_radius_vs_tails}) but it clearly aggravates non-radiative recombination losses, as evidenced by a particularly low PL intensity in CBTS and CSTS (Fig.~\ref{fig:correlations}(c)). 
We speculate that interstitial defects, which generally have high formation energies in kesterite or stannite materials,~\cite{Chen2013,Yuan2015} may be abundant in CBTS and CSTS due to the characteristics of the $P3_1$ structure (Fig.~\ref{fig:structures}(c)). 
Specifically, the eight sulfur atoms surrounding the II$^{2+}$ cation in the $P3_1$ structure form square antiprisms. 
The squared faces of the antiprisms are rather distant from one another, leaving an open "cage" where an interstitial atom can be accommodated without a large perturbation of the crystal structure. 
In CBTS, a S interstitial located between two antiprisms would be $\sim$3.20~\AA ~away from eight other sulfur atoms, which is not much closer than the equilibrium distance between sulfur atoms in the $P3_1$ structure of CBTS (3.52~\AA). 
It is also important to note that defects with $\sim$600~meV ionization energy have been identified in CBTS and CSTS by PL characterization in this study (Fig.~\ref{fig:defect_levels}). With some rare exceptions~\cite{Grossberg2014} such deep defects are usually not detected by PL in materials with the kesterite structure. Hence, we conclude that deep defects and non-radiative recombination are a serious issue in the $P3_1$-structured CBTS and CSTS compounds.

\section{Conclusion}
To better understand the fundamental loss mechanisms of kesterite-inspired photovoltaic materials, we investigated the chemical trends in the defect properties of the Cu$_2$-II-Sn-S$_4$ (CXTS) series by a combination of experimental techniques and first-principles calculations. Three main conclusions were reached. First, the dominant shallow defects in CBTS and CSTS are Cu vacancies (acceptors) and Cu interstitials (donors) instead of the Cu$_\mathrm{II}$ and II$_\mathrm{Cu}$ antisites that are prominent in CZTS. Second, band tailing in the CXTS series decreases gradually as the size mismatch between the II$^{2+}$ cation and the other cations increases. Third, mitigation of band tailing does not imply mitigation of non-radiative recombination rates. On the contrary, deep defects are more prominent and non-radiative recombination losses more severe in CBTS and CSTS than in CZTS, despite their much less pronounced band tails.

Then, we took a broader look at the family of kesterite-inspired semiconductors and suggested that deep Urbach tails and shallow Gaussian tails may have fundamentally different origins in this class of materials. Urbach tails are correlated with the calculated band gap narrowing caused by (2I$_\mathrm{II}$+IV$_\mathrm{II}$) defect clusters. Gaussian tails are correlated with the energy difference between the kesterite and stannite polymorphs in the materials having kesterite as the lowest-energy structure. Noticing that the transformation from kesterite to stannite is equivalent to the formation of (I$_\mathrm{II}$+II$_\mathrm{I}$)$_{\perp}$ defect clusters across cationic planes, we suggest that Gaussian tails may be caused by these out-of-plane clusters. Unlike the (I$_\mathrm{II}$+II$_\mathrm{I}$)$_{\parallel}$ clusters responsible for the well-known cation disorder in kesterite, the different arrangement of (I$_\mathrm{II}$+II$_\mathrm{I}$)$_{\perp}$ clusters implies that they do not cause disorder in the I-II plane, which is the experimentally available measure of the (I$_\mathrm{II}$+II$_\mathrm{I}$) defect concentation.
This subtle difference can explain why various studies have found a lack of correlation between cation disorder and band tails in kesterite. Our results provide clear criteria for selecting tail-state-free photovoltaic absorbers in the vast I$_2$-II-IV-VI$_4$ chemical space.

\section*{Acknowledgements}
This project has received funding from the European Union’s Horizon 2020 research and innovation programme under the Marie Sk\l odowska-Curie grant agreement No 840751, and from VILLUM Fonden (grant no. 9455). The Center for Nanostructured Graphene is sponsored by the Danish National Research Foundation (Project No. DNRF103). The PL imaging setup has been partly funded by the IDUN Center of Excellence funded by the Danish National Research Foundation (project no. DNRF122) and VILLUM Fonden (grant no. 9301). This research has also been funded by the EU Horizon2020 Framework (STARCELL, Grant No. 720907). We are grateful to the UK Materials and Molecular Modelling Hub for computational resources, which is partially funded by EPSRC (EP/P020194/1). Via our membership of the UK’s HEC Materials Chemistry Consortium, which is funded by EPSRC (EP/L000202), this work used the ARCHER UK National Supercomputing Service (http://www.archer.ac.uk).

%\section*{Text in the Supplementary material}
%\balance  % balances the last columns

% REFXRENCES
\section{References}
\bibliography{library}

\newpage
\clearpage

% Resets the figure counter and labels ssupplementary figures by S#
%\renewcommand{\thefigure}{S\arabic{figure}}
%\setcounter{figure}{0}

%\balance  % balances the last columns
\onecolumngrid

\section*{SUPPORTING INFORMATION}

\subsection*{The special case of band gap narrowing in AZTSe}
In Fig.~8(a) of the main article, the calculated band gap narrowing due to the 2(I$_{\mathrm{II}}$ + IV$_{\mathrm{II}}$) defect cluster is plotted on the x-axis, based on the values calculated in Refs.~\citenum{Chen2013,Yuan2015}. The AZTSe data point is, however, a special case that needs further clarification. First, calculations of defect clusters are only available for AZTS and not for AZTSe and, second, the 2(Ag$_{\mathrm{Zn}}$ + Sn$_{\mathrm{Zn}}$) cluster in AZTS has a very high formation energy.~\cite{Yuan2015} As a worst-case scenario, we considered the defect cluster with the largest calculated band gap narrowing among low-formation-energy defects in AZTS, which is (Sn$_{\mathrm{Zn}}$ + Zn$_{\mathrm{Sn}}$). To extrapolate its band gap narrowing effect from AZTS to AZTSe we simply followed the trend of the (Sn$_{\mathrm{Zn}}$ + Zn$_{\mathrm{Sn}}$) defect from CZTS to CZTSe, as shown in Ref.~\citenum{Chen2013}. We consider this an acceptable approximation since all four compounds have the same crystal structure (kesterite) and exhibit the same band gap trend, i.e., a band gap decrease by $\sim$0.5~eV when replacing S with Se.

When going from CZTS to CZTSe, the valence band upshift is reduced by a factor $\sim$2, and the conduction band downshift is reduced by a factor of $\sim$~4.\cite{Chen2013} Using these correction factors, the calculated band gap narrowing of AZTSe is extrapolated as 70~meV, which is the value plotted on the x-axis of Fig.~8(a) in the main article for the AZTSe data point. Note that the trend identified in Fig.~8(a) is still valid if a different defect cluster is chosen for AZTSe, because all the other calculated defect clusters narrow the band gap to a smaller extent than the (Sn$_{\mathrm{Zn}}$ + Zn$_{\mathrm{Sn}}$) cluster.~\cite{Yuan2015}

\renewcommand{\thefigure}{S1}
\begin{figure*}[t!]
\centering%
\includegraphics[width=\textwidth]{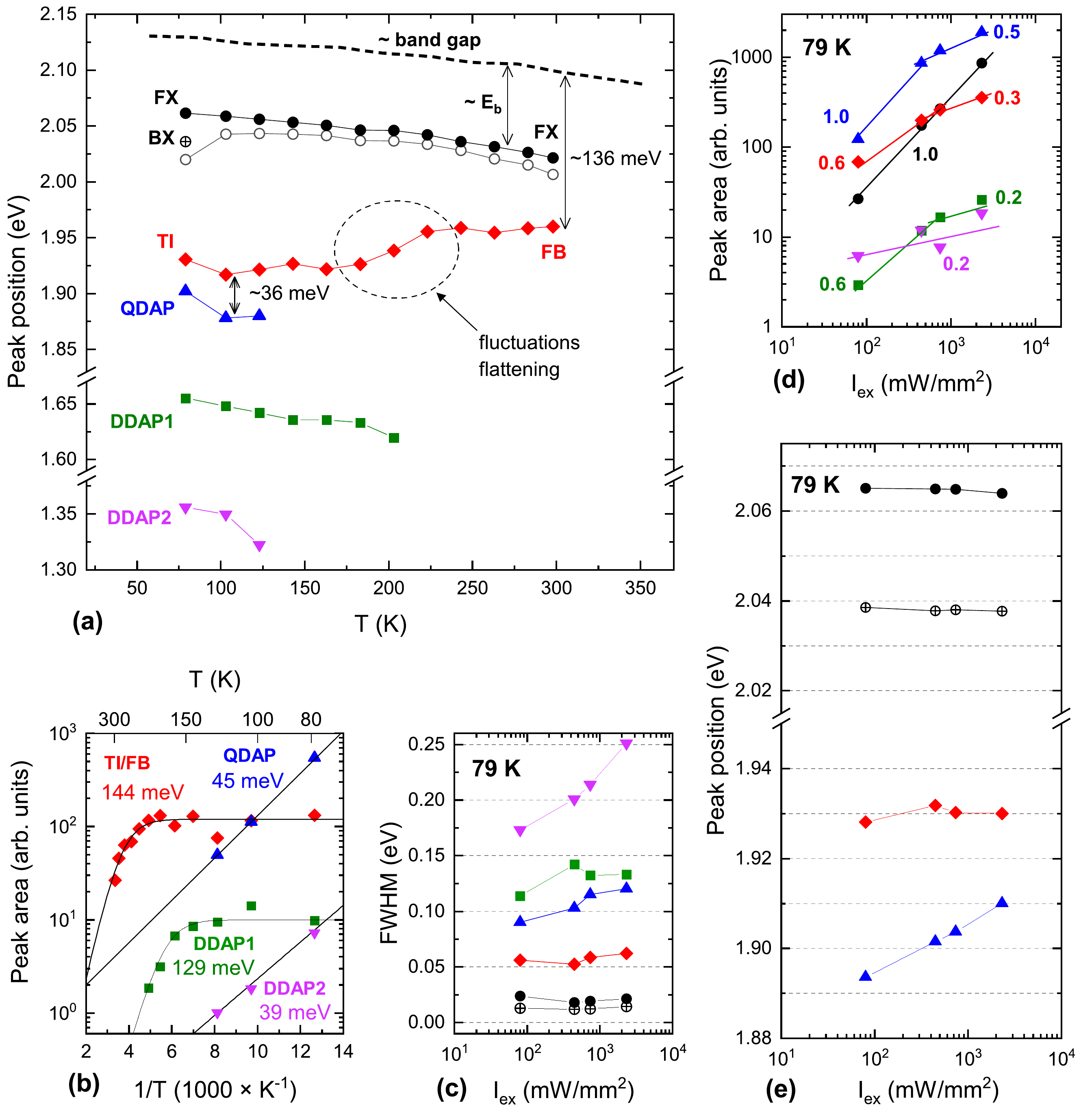}
\caption{Integrated area, position, and full-width at half-maximum (FWHM) of the PL peaks identified in a CBTS film as a function of temperature and excitation intensity. The parameters of each peak are obtained by least-squares fitting with a single Gaussian peak. (a): Peak position versus temperature. Each peak is labeled with the type of PL transition responsible for it. The exciton binding energy estimated from the hydrogen model, and the consequent position of the band gap energy as a function of temperature, are indicated. The temperature range corresponding to flattening of band edge fluctuations due to state filling is also indicated. (b): Arrhenius plots of PL peak areas as a function of temperature The type of transition and the fitted activation energy is indicated for each peak. (c): FWHM versus excitation intensity at 79~K. The color- and marker scheme is the same as in the previous subfigures. (d): Peak area versus excitation intensity at 79~K. The power law coefficients $k$ under low- and high excitation are shown. (e): Peak position versus excitation intensity at 79~K. The intensity of the DDAP1 and DDAP2 peaks is too low for the small excitation-dependent positions shifts to be determined reliably.}
\label{fig:CBTS_PL_data}
\end{figure*}

\renewcommand{\thefigure}{S2}
\begin{figure*}[t!]
\centering%
\includegraphics[width=\textwidth]{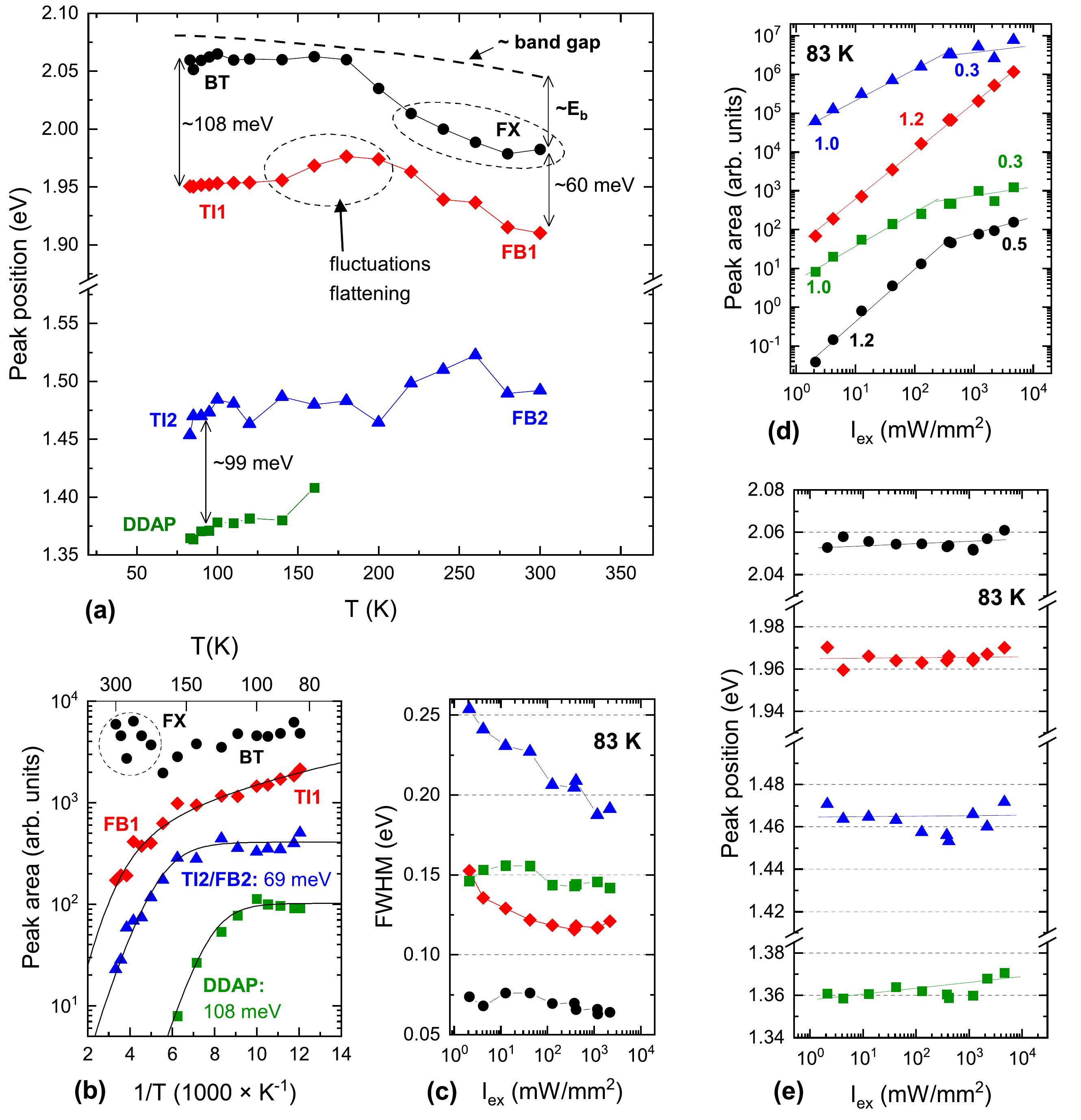}
\caption{Integrated area, position, and full-width at half-maximum (FWHM) of the PL peaks identified in a CSTS film as a function of temperature and excitation intensity. The parameters of each peak are obtained by least-squares fitting with a single Gaussian peak. (a): Peak position versus temperature. Each peak is labeled with the type of PL transition responsible for it. The exciton binding energy estimated from the hydrogen model, and the consequent position of the band gap energy as a function of temperature, are indicated. The temperature range corresponding to flattening of band edge fluctuations due to state filling is also indicated. (b): Arrhenius plots of PL peak areas as a function of temperature The type of transition and the fitted activation energy is indicated for each peak. (c): FWHM versus excitation intensity at 83~K. The color- and marker scheme is the same as in the previous subfigures. (d): Peak area versus excitation intensity at 83~K. The power law coefficients $k$ under low- and high excitation are shown. (e): Peak position versus excitation intensity at 83~K.}
\label{fig:CSTS_PL_data}
\end{figure*}

%\clearpage
%%\bibliography{library}
%%\bibliographystyle{iopart-num}
%
%\providecommand{\newblock}{}
%\begin{thebibliography}{1}
%\expandafter\ifx\csname url\endcsname\relax
%  \def\url#1{{\tt #1}}\fi
%\expandafter\ifx\csname urlprefix\endcsname\relax\def\urlprefix{URL }\fi
%\providecommand{\eprint}[2][]{\url{#2}}
%% Bibliography created with iopart-num v2.1
%% /biblio/bibtex/contrib/iopart-num
%
%\bibitem{Chen2013}
%Chen S, Walsh A, Gong X~G and Wei S~H 2013 {\em Advanced Materials\/} {\bf 25}
%  1522--1539
%
%\bibitem{Yuan2015}
%Yuan Z~K, Chen S, Xiang H, Gong X~G, Walsh A, Park J~S, Repins I and Wei S~H
%  2015 {\em Advanced Functional Materials\/} {\bf 25} 6733--6743
%
%\end{thebibliography}

\end{document}